\documentclass[12pt]{article}

\usepackage{amsmath,amssymb,amsthm}
\usepackage{setspace}
\usepackage[dvipsnames]{xcolor}
\usepackage{blindtext}
\usepackage[margin=1in]{geometry}
\usepackage{natbib}
\usepackage{graphicx}
\usepackage{booktabs}
\usepackage{multirow}
\usepackage{mathrsfs}
\usepackage{bm}
\usepackage[shortlabels]{enumitem}
\usepackage{pdfpages}

\newcommand{\mythanks}[1]{\thanks{\parbox[t]{0.95\textwidth}{#1}}}

\definecolor{mylinkcolor}{HTML}{e83746}
\colorlet{myurlcolor}{violet}
\colorlet{mycitecolor}{Aquamarine}

\newcommand{\Normal}{\operatorname{Normal}}
\newcommand{\Reals}{\mathbb R}
\newcommand{\Identity}{\mathrm I}
\newcommand{\iid}{\stackrel{\textnormal{iid}}{\sim}}
\newcommand{\trace}{\operatorname{tr}}
\newcommand{\Bernoulli}{\operatorname{Bernoulli}}
\newcommand{\sA}{\mathscr A}
\newcommand{\GP}{\operatorname{GP}}

\newcommand{\E}{\mathbb E}
\newcommand{\diag}{\operatorname{diag}}
\newcommand{\asim}{\stackrel{{}_{\bullet}}{\sim}}
\newcommand{\Var}{\operatorname{Var}}
\newcommand{\Cov}{\operatorname{Cov}}
\newcommand{\Ell}{\mathscr L}
\newcommand{\Uniform}{\operatorname{Uniform}}
\newcommand{\Span}{\operatorname{span}}
\newcommand{\indep}{\stackrel{\textnormal{indep}}{\sim}}
\newcommand{\sM}{\mathscr M}

\newcommand{\bX}{\bm X}
\newcommand{\bY}{\bm Y}
\newcommand{\bA}{\bm A}

\newcommand{\bE}{\bm E}

\newcommand{\uS}{\underline{S}}
\newcommand{\dfrak}{\mathfrak d}

\newtheorem{proposition}{Proposition}

\newtheorem{principle}{Principle}
\newtheorem{theorem}{Theorem}
\theoremstyle{remark}
\newtheorem{remark}{Remark}
\theoremstyle{definition}

\usepackage[colorlinks = true, citecolor = NavyBlue, linkcolor = mylinkcolor]{hyperref}

\begin{document}

\author{Antonio R. Linero\mythanks{Department of Statistics and Data Sciences, University of Texas at Austin, email: \href{mailto:antonio.linero@austin.utexas.edu}{antonio.linero@austin.utexas.edu}}}
\date{}
\title{In Nonparametric and High-Dimensional Models, Bayesian Ignorability is an Informative Prior}

\maketitle

\begin{abstract}
  In problems with large amounts of missing data one must model two distinct data generating processes: the outcome process which generates the response and the missing data mechanism which determines the data we observe. Under the \emph{ignorability} condition of \citet{rubin1976}, however, likelihood-based inference for the outcome process does not depend on the missing data mechanism so that only the former needs to be estimated; partially because of this simplification, ignorability is often used as a baseline assumption. We study the implications of Bayesian ignorability in the presence of high-dimensional nuisance parameters and argue that ignorability is typically incompatible with sensible prior beliefs about the amount of selection bias. We show that, for many problems, ignorability directly implies that the prior on the selection bias is tightly concentrated around zero. This is demonstrated on several models of practical interest, and the effect of ignorability on the posterior distribution is characterized for high-dimensional linear models with a ridge regression prior. We then show both how to build high-dimensional models which encode sensible beliefs about the selection bias and also show that under certain narrow circumstances ignorability is less problematic.
\end{abstract}

\doublespacing

\section{Introduction}

Dealing with missing data is a fundamental problem in data analysis; for example, missingness complicates inference in clinical trials \citep{national2010The} and is inherent in the potential outcomes framework for causal inference \citep{rubin2005causal}. A common starting point for addressing missingness is to assume that the mechanism which generated the missingness is \emph{ignorable} \citep{rubin1976}. Ignorability allows likelihood-based inference to proceed without modeling the missing data mechanism, which can greatly simplify an analysis.

In this paper we consider the Bayesian approach to account for missingness. For generality, we consider a potential outcome $Y_i(a)$ for some exposure level $a \in \sA$ such that we observe both the exposure level $A_i$ and its associated potential outcome $Y_i(A_i)$ ($Y_i(a)$ is regarded as missing for all $a \ne A_i)$. Let $X_i$ be a vector of confounders which are predictive of both $A_i$ and $Y_i(a)$. By defining $Y_i(1)$ as the outcome of interest, this framework subsumes the standard missing data problem, where $A_i$ is now a missing data indicator such that we observe the outcome when $A_i = 1$. We say that the \emph{missing data mechanism} $f_\phi(A_i \mid X_i)$ is Bayesian-ignorable, or simply ignorable, if the following conditions hold.
\begin{enumerate}[label={IG.\arabic*}]
\item
  The potential outcomes $\{Y_i(a) : a \in \sA\}$ are conditionally independent of $A_i$ given $X_i$.
  \label{ig1}
\item
  The parameters $\beta$ and $\phi$ are a-priori independent, where $\beta$ parameterizes the model for the potential outcomes and $\phi$ parameterizes the missing data mechanism. That is, the prior factors as $\pi(\beta,\phi) = \pi_\beta(\beta) \, \pi_\phi(\phi)$. \label{ig3}
\end{enumerate}
Condition \ref{ig1} constrains the data generating mechanism and is a (type of) \emph{missing at random} (MAR) assumption \citep{seaman2013meant}; in the causal inference literature, assumptions like \ref{ig1} are sometimes themselves referred to as strong ignorability assumptions \citep{rosenbaum1983central,imai2010general}, and it is sometimes conflated with ignorability in missing data sense of \citet{rubin1976} as well \citep[see][for a through discussion of MAR and ignorability]{seaman2013meant}. Condition \ref{ig3}, which constrains the prior, is also key to ignorability: it guarantees that the posterior distribution of $\beta$ given the observed data is proportional to $\pi_\beta(\beta) \ \prod_{i} f_\beta\{Y_i(A_i) \mid X_i\}$,
which does not depend on the missing data mechanism. Without \ref{ig3} we are still obligated to model the missing data mechanism even when the missing data is MAR, as $A_i$ provides information about $\beta$ through $\phi$.

It has been argued before that, from a Frequentist perspective, \ref{ig3} is highly problematic in high-dimensional problems \citep{robins1997toward}. We complement this Frequentist view and study the implications of \ref{ig3} from a Bayesian perspective. In particular we will argue that, while \ref{ig3} is seemingly innocuous, it actually is highly informative about the selection bias in high-dimensional problems to the degree that the data has no reasonable chance at overcoming the prior. We refer to this as \emph{prior dogmatism} about the selection bias. We make the following three points.
\begin{enumerate}[leftmargin = *]
\item
  Priors which impose \ref{ig3} are dogmatic about the amount of selection bias. This is particularly true in models which require the use of informative priors, such as high-dimensional or Bayesian nonparametric models. We conclude that \ref{ig3} does not reflect substantive prior knowledge in most cases; in the case of a causal ridge regression model, we are able to quantify these problems explicitly using random matrix theory (see \citealp{dobriban2018high} and \citealp{dicker2016ridge} for related applications of random matrix theory). 
\item
  By understanding this induced prior on the selection bias, we are able to identify several highly effective ways of correcting this problem and unify several approaches proposed in the Bayesian causal inference literature which were not motivated by Bayesian considerations. Our remedies take the form of propensity score adjustments, which have typically been recommended in applied Bayesian analysis on the grounds of pragmatism and robustness \citep[see, e.g.,][]{rubin1985use} rather than subjective Bayesian principles.
\item
  We study some relatively narrow settings in which prior dogmatism does not occur, even in high dimensional problems. For example, strong dependence structure in $X_i$ can act as a shield against dogmatism; in the case of causal ridge regression, we again use random matrix theory to quantify this behavior. Despite this, we find little payoff for failing to correct for dogmatism in these settings.
\end{enumerate}

\begin{remark}
  While we will consider the Frequentist properties of the posterior distribution, our goal at the outset is not to construct priors specifically to attain ideal Frequentist properties. It is often quite easy to construct priors which are doubly robust and attain some semiparametric efficiency bound if that is our goal from the outset, and various complete class theorems \citep[][Chapter 8]{robert2007bayesian} suggest that we can usually construct \emph{some} Bayes estimator which is at-least-as-good as any given Frequentist estimator. Rather, we (i) show that priors of the form \ref{ig3} are inherently dangerous in purely-Bayesian terms, (ii) explain in which situations the problem is most acute, and (iii) use dogmatism to show where corrections are needed. 
\end{remark}

\subsection{Notation}

We let $Y_i(\cdot) \in \Reals$ denote an outcome, $X_i \in \Reals^P$ a covariate/confounder, and $A_i \in \Reals$ a treatment/missing data indicator for $i = 1,\ldots,N$. When considering causal inference problems, we define $Y_i = Y_i(A_i)$ to be the observed outcome; when $A_i$ is a missingness indicator, we instead define $Y_i = Y_i(1)$ so that $Y_i$ is missing when $A_i \ne 1$. We set $\bY = (Y_1,\ldots,Y_N)^\top$, $\bA = (A_1, \ldots, A_N)$, and let $\bX$ denote an $N \times P$ matrix obtained by stacking the row vectors $X_i^\top$. Let $\beta$ parameterize the distribution of $[Y_i(\cdot) \mid X_i]$, let $\phi$ parameterize the distribution of $[A_i \mid X_i]$, and let $\theta = (\beta,\phi)$. We invoke \ref{ig1} throughout. 
We let $\E_\theta(\cdot)$ denote the expectation operator conditional on $\theta$. If the subscript $\theta$ is omitted then $\E(\cdot)$ is the expectation operator with respect to a prior distribution on $\theta$, e.g., $\E(Y_i) = \int \E_\theta(Y_i) \, \pi(\theta) \ d\theta$. We use the Big-O notation $W = O_p(V)$ to mean that $|W|/|V|$ is bounded in probability as $P\to\infty$ (with the dependence of $W$ and $V$ on $P$ suppressed).

\subsection{Three Illustrative Problems}
\label{sec:three}

We consider three problems to illustrate the existence of dogmatism and how to correct for it. The first two concern causal inference with a continuous exposure, while the third is a missing data problem. We assume $X_i \sim \Normal(0, \Sigma)$ for some $\Sigma \in \Reals^{P \times P}$ to simplify our analysis. All proofs are deferred to the Supplementary Material.

\begin{paragraph}{Ridge regression in causal inference}
  Let $Y_i(a)$ denote  the outcome observed when individual $i$ receives the level $a$ of some continuous exposure and let $A_i$ denote the value of the exposure which is actually received. We observe $(A_i, Y_i)$ where $Y_i = Y_i(A_i)$. We posit the linear models $Y_i(a) = X_i^\top \beta + \gamma \, a + \epsilon_i(a)$ and $A_i = X_i^\top\phi + \nu_i$ with $\epsilon_i(a) \sim \Normal(0, \sigma^2_y)$ and $\nu_i \sim \Normal(0, \sigma^2_a)$. The Bayesian ridge regression prior, which satisfies \ref{ig3}, takes $\beta \sim \Normal(0, \tau^2_\beta \ \Identity)$ and $\phi \sim \Normal(0, \tau^2_\phi\ \Identity)$. The parameter of inference is the mean response at a given exposure $\E_\theta\{Y_i(a)\} = \gamma \, a$. We assume that $X_i$ is high-dimensional in the sense that $P$ is potentially larger than $N$.
\end{paragraph}

\begin{paragraph}{Sparsity priors in causal inference}
  This is the same problem as the ridge regression problem, except that $\beta$ and $\phi$ are sparse. We consider independent spike-and-slab priors for the coefficients, i.e., $\beta_j \iid (1 - p_\beta) \, \delta_0 + p_\beta \, \Normal(0, \tau^2_\beta)$ and $\phi_j \iid (1 - p_\phi) \, \delta_0 + p_\phi \, \Normal(0,\tau^2_\phi)$ where $\delta_0$ denotes a point-mass distribution at $0$.
\end{paragraph}

\begin{paragraph}{Semiparametric regression with missing data}
  An outcome $Y_i = Y_i(1)$ is observed if $A_i = 1$ and missing if $A_i = 0$, and our goal is to estimate $\E_\theta(Y_i)$. 
  We consider the model $Y_i = \beta(X_i) + \epsilon_i$ with $\epsilon_i \sim \Normal(0, \sigma^2_y)$ and $A_i \sim \Bernoulli\{\phi(X_i)\}$. We assume that $\beta(\cdot)$ has a \emph{Gaussian process} prior \citep{rasmussen2005gaussian} with covariance function $\kappa(\cdot,\cdot)$, written $\GP(0, \kappa)$.
\end{paragraph}



\section{The Induced Prior on the Selection Bias}

The fundamental difficulty with missingness is \emph{selection bias}. When estimating $\E_\theta\{Y_i(a)\}$ this amounts to the fact that $\Delta(a) = \E_\theta\{Y_i(a) \mid A_i = a\} - \E_\theta\{Y_i(a)\} \ne 0$. That the selection bias parameter $\Delta(a)$ is non-zero is the only feature of the problem which makes estimation of $\E_\theta\{Y_i(a)\}$ non-trivial, as otherwise we could ignore the covariates $X_i$ and directly estimate $\E_\theta\{Y_i(a)\}$ by estimating $\E_\theta\{Y_i(a) \mid A_i = a\}$ nonparametrically. The following proposition gives an expression for $\Delta$ in each of our problems.

\begin{proposition}
  \label{prop:ridge-bias}
  In the ridge and spike-and-slab regression problems, the selection bias is given by
  \begin{math}
    \Delta(a)
    = 
    a \,\dfrac{\phi^\top \Sigma \beta}{\sigma^2_a + \phi^\top\Sigma\phi}
    =
    a \,
    \dfrac{\sum_j \lambda_j \, W_j \, Z_j}
         {\sigma^2_a + \sum_j \lambda_j \, Z_j^2}
  \end{math}
  where $W = \Gamma^\top \beta$, $Z = \Gamma^\top \phi$, and $\Sigma = \Gamma \Lambda \Gamma^\top$ is the spectral decomposition of $\Sigma$ with $\Lambda = \diag(\lambda_1, \ldots, \lambda_P)$. In the semiparametric regression problem with missing data, we instead have
  \begin{math}
      \Delta \equiv \Delta(1) = 
      {\Cov_\theta\{\beta(X_i), \phi(X_i)\}}/{\E_\theta\{\phi(X_i)\}}.
  \end{math}
\end{proposition}

Given the importance $\Delta$ and the working assumption that selection bias is non-negligible, one would hope that the prior distribution of $\Delta$ is relatively diffuse. Using Proposition~\ref{prop:ridge-bias} we can gain insight into how the prior on the selection bias changes as the dimension $P$ increases. For example, for the ridge regression problem we have the following result.

\begin{proposition}
  \label{prop:ridge-clt}
  Assume the setup of Proposition~\ref{prop:ridge-bias} for the ridge regression problem and suppose $\beta \sim \Normal(0, \tau^2_\beta \, \Identity)$ and $\phi \sim \Normal(0, \tau_\phi^2 \, \Identity)$ independently. Assume $\frac{1}{P} \sum_{j=1}^P \lambda_j^k$ converges to a positive constant as $P \to \infty$ for $k = 1, 2, 2 + \epsilon$ for some $\epsilon$, and let $\widetilde \lambda$ and $\bar \lambda^2$ be the limits with $k = 1,2$. Then 
  \begin{math}
    \Delta(a) 
    \asim
    \Normal(
    0, c/P
    )
  \end{math}
  where $c = a^2 \, (\tau^2_\beta/\tau^2_\phi) \, (\bar \lambda^2 /  \widetilde\lambda^2)$.
\end{proposition}

We will return to the conditions on $\Sigma$ (which are moment conditions on the spectral distribution of $\Sigma$) later and focus on the conclusion $\Delta(a) \asim \Normal(0, c/P)$ for some constant $c$. If selection bias is a-priori of concern for us then it seems unwise to posit a $\Normal(0, c/P)$ prior for it when $P$ is large. This behavior becomes even more suspect when one considers that the definition of $\Delta(a)$ is completely free of the $X_i$'s, and that logically the act of measuring additional covariates should not change our beliefs about $\Delta(a)$. In Section~\ref{sec:asymptotics} we follow up on the inferential consequences of this.


At a high level, the source of the problem in our three illustrative examples is the following well-known phenomenon which we refer to as the \emph{orthogonality principle}.

\begin{principle}[The Orthogonality Principle]
   Let $\widetilde \beta$ and $\widetilde \phi$ be random unit vectors with mean $0$ taking values in some high/infinite dimensional Hilbert space $\mathcal H$ with inner product $\langle \cdot, \cdot \rangle$. Then, if $\widetilde \phi$ and $\widetilde\beta$ are independent and there is no other special structure in the problem, with high probability we have $\langle \widetilde\beta, \widetilde\phi\rangle \approx 0$. 
\end{principle}

For a concrete example, by the law of large numbers and the central limit theorem, if $\beta_j, \phi_j \iid \Normal(0,1)$ then $\beta^\top\phi/\sqrt P \to \Normal(0,1)$ but $\|\beta\| \, \|\phi\|/P \to 1$ so that $\langle \beta / \|\beta\|, \phi / \|\phi\|\rangle = O_p(P^{-1/2})$ with respect to the Euclidean inner product. The statement of the orthogonality principle is intentionally vague as to what it means for the dimension to be ``high,'' what $\approx$ means, and what constitutes ``special structure.'' Nevertheless, it provides immediate intuition for what to expect: unless one has reason to believe otherwise, expect high-dimensional unit vectors to be nearly orthogonal.

The orthogonality principle becomes important when $P$ is large (or in nonparametric problems) because $\Delta$ is quantifiable in terms of $\langle \beta, \phi \rangle$ for some suitable inner product (c.f. Proposition~\ref{prop:ridge-bias}). If \ref{ig3} holds then the orthogonality principle immediately suggests $\langle \beta, \phi \rangle \approx 0$ with high probability, implying that our prior is dogmatic about the selection bias.

\subsection{Asymptotics for High-Dimensional Ridge Regression}
\label{sec:asymptotics}

While the dogmatism implied by Proposition~\ref{prop:ridge-clt} is troubling, one might hope that the informative prior on $\Delta$ is a theoretical curiosity which is nevertheless swamped by the data. By analogy, a strict prior analysis of the flat prior $\mu \sim \Normal(0, 10^{100})$ in the normal model $Y_i \iid \Normal(\mu, 1)$ would similarly suggest that we ought to be concerned about the implication of the prior on the magnitude of $\mu$; instead the data quickly swamps the diffuse prior to produce a sensible posterior $\mu \approx \Normal(\bar Y, 1/N)$. That is, in terms of consequences for inference, what matters is the impact of \ref{ig3} on the posterior rather than the prior.

We now show that this hopeful scenario is not borne out, and that the prior concentration on $\Delta$ leads to heavily biased inferences if $P$ grows sufficiently quickly with $N$. We summarize our main results as follows:
\begin{itemize}[leftmargin = *]
    \item In the regime $P/N \to r$ for some $r \in (0,\infty)$ (i.e., $P$ grows at the same rate as $N$), the Bayes estimator which takes a flat prior on $\gamma$ and a Gaussian prior $\beta \sim \Normal(0,\tau^2 \, P^{-1} \, \Identity)$ is heavily biased. Specifically, when selection bias is present through the auxiliary covariate $\widehat A_i = X_i^\top \phi$, the Bayes estimate will have bias of order $\Delta(1)$.
    \item In some sense, the setting $\Sigma = \Identity$ is inherently difficult, and the problem is generally easier when the components of $X_i$ are highly correlated. We return to this point in Section~\ref{sec:when}.
\end{itemize}

We make two sets of assumptions. The first (high-dimensional asymptotics, or HDA) is used to describe the distribution of the $X_i$'s as $N \to \infty$. The second (random effects model, or REM) describes a particular random effects model for the regression coefficients. This framework modifies the framework of \citet{dobriban2018high} so that it is suitable for our aims.




\begin{enumerate}[label={HDA.\arabic*}]
  \item The covariates are multivariate normal with $X_i \sim \Normal(0,\Sigma)$.\label{hda1}
  \item As $N \to \infty$ we have $P/N \to r$ for some $r \in (0,\infty)$.\label{hda2}
  \item The spectral distribution $\sum_{p=1}^P \delta_{\lambda_p} / P$ associated to $\Sigma$ converges to some limiting distribution $H$ on $[0,\infty)$, where $\lambda_1,\ldots,\lambda_P$ are the eigenvalues of $\Sigma$ and $\delta_\lambda$ denotes a point-mass distribution at $\lambda$.\label{hda3}
\end{enumerate}

HDA is a standard assumption for understanding the case where $P$ grows like $N$, though \ref{hda1} may be replaced with a moment condition on $X_i$. \ref{hda3} allows us to use results from random matrix theory to compute $\lim_{P\to\infty}\trace\{(\bX^\top\bX + N\lambda\,\Identity)^{-k}\}$ for $k \in \mathbb N $. Under HDA, the empirical distribution of the eigenvalues of $\uS = \bX\bX^\top/N$, namely $\widehat F(dx) =\sum_{i=1}^N \delta_{\widehat \lambda_i}/N$, converges to a distribution $F(dx)$ called the \emph{empirical spectral distribution}.


Next, we describe a random effects model (REM) for $\beta$ and $\phi$ we will base our analysis on. Similar models have been used to study both the prediction risk and minimax-optimality of ridge regression \citep{dobriban2018high, dicker2016ridge}. REM is a fruitful assumption for us as it allows exact formulas for the bias to be derived which are free of the particular values of $\beta$ and $\phi$.

\begin{enumerate}[label={REM.\arabic*}]
    \item The coefficient vector $\phi$ is randomly sampled as $\phi \sim \Normal(0, \tau^2 \, P^{-1} \, \Identity)$.\label{rrc1}
    \item The coefficient vector $\beta$ is randomly sampled as $\beta \sim \Normal(\omega_0 \, \phi, \tau^2 \, P^{-1} \, \Identity)$.\label{rrc2}
    \item Given $\beta$ and $\phi$, $Y_i \sim \Normal(X_i^\top\beta + A_i \, \gamma_0, 1)$ and $A_i \sim \Normal(X_i^\top\phi, 1)$.
\end{enumerate}

To motivate \ref{rrc2}, note that it is equivalent to setting $Y_i \sim \Normal(X_i^\top b + \omega_0 \, \widehat A_i + \gamma_0 \, A_i, 1)$ where $\widehat A_i = X_i^\top\phi = \E(A_i \mid X_i, \phi)$ and $b \sim \Normal(0,\tau^2 \, P^{-1} \, \Identity)$. \ref{rrc2} allows for non-negligible selection bias to enter the model, and priors based on this parameterization have been used to account for selection bias by other researchers \citep{zigler2013model,hahn2018regularization}. The parameter $\omega_0$ is intimately connected to the selection bias.

\begin{proposition}
  \label{prop:omega}
  Suppose that HDA and REM hold and that $\Sigma$ satisfies the conditions of Proposition~\ref{prop:ridge-clt}. Then
  \begin{math}
      \Delta(1)
    \to 
    \omega_0 \frac{\tau^2 \, \widetilde\lambda}{1 + \tau^2 \, \widetilde \lambda}
  \end{math}
  in probability as $P \to \infty$.
\end{proposition}

Theorem~\ref{thm:ridge} explicitly computes the bias of the ridge regression estimator under \ref{ig3} when the prior $\beta \sim \Normal(0, N^{-1} \lambda^{-1} \Identity)$ is used, i.e., when we apply the usual ridge regression estimator. We sketch a proof of Theorem~\ref{thm:ridge} and verify it numerically in the Supplementary Material.


\begin{theorem}
    \label{thm:ridge}
    Suppose HDA and REM hold. Let $(\widetilde\gamma, \widetilde \beta^\top)^\top$ denote the Bayes estimate of $(\gamma, \beta^\top)^\top$ under a prior which takes $\beta \sim \Normal(0, N^{-1} \, \lambda^{-1} \Identity)$ and places a flat prior on $\gamma$ under \ref{ig3}. Then the asymptotic bias of $\widetilde\gamma$ is given by
    \begin{align}
        \label{eq:bias}
        \lim_{N,P\to\infty} \E(\widetilde\gamma - \gamma_0)
        =
        \frac{\omega_0 \int x/(x+\lambda) \ F(dx)}{\int (x+\eta)/(x+\lambda) \ F(dx)}
        =
        \omega_0 \times 
        \frac{1 - \lambda \, v(-\lambda)}{1 - (\lambda - \eta) \, v(-\lambda)}
    \end{align}
    where $v(z) = \int_0^\infty \frac{F(dx)}{x-z}$ is the Stieltjes transform of $F(dx)$ and $\eta = r / \tau^2$.
    
\end{theorem}



\begin{figure}[t]
    \centering
    \includegraphics[width=.9\textwidth]{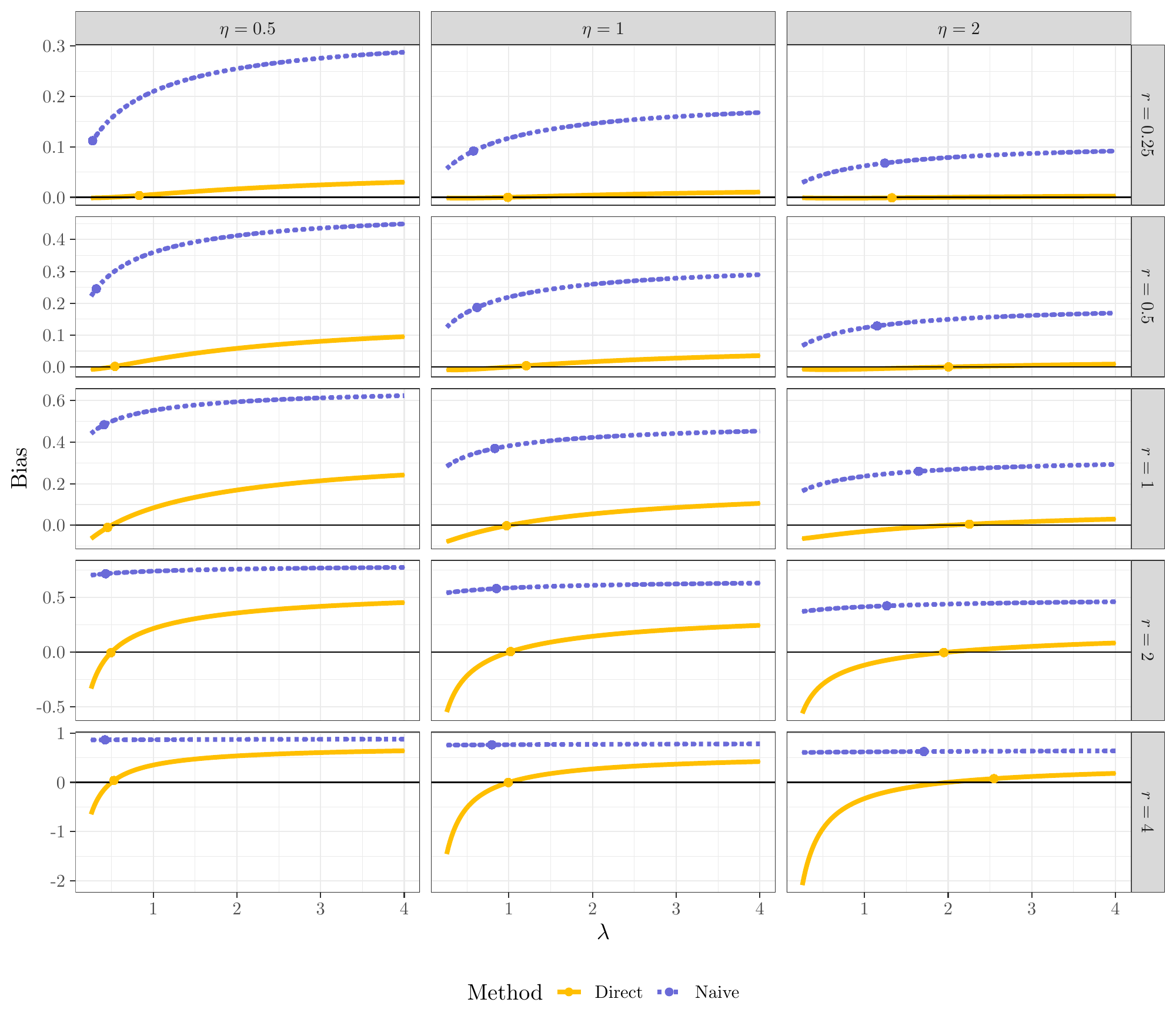}
    \caption{Comparison of the bias of naive ridge regression (dashed, blue) to the direct Z-prior (solid, orange) of Section~\ref{sec:direct} for different values of $\eta$ and $r$ with $\omega_0 \equiv 1$. Estimated values of $\lambda$ based on a single simulated dataset for each combination of $\eta$ and $r$ are given by the points.}
    \label{fig:bias_compare}
\end{figure}

Ideally we would like the bias to be close to $0$ for moderate-to-large values of $\lambda$ so that we have both small variance and bias; the approach outlined in Section~\ref{sec:direct} \emph{does} accomplish this goal for a properly-chosen $\lambda$. Figure~\ref{fig:bias_compare} contrasts this alternative method with standard ridge regression when $\Sigma = \Identity$ and we see that the bias is quite large for ridge regression unless $\lambda$ is close to $0$ and $r \le 1$; this latter case corresponds to OLS, which (while unbiased) defeats the purpose of using ridge regression.

A qualitative observation based on \eqref{eq:bias} is that smaller bias is obtained when most of the eigenvalues of $\uS$ are small. For example, unbiasedness is possible if $F(dx)$ assigns \emph{any} mass to $0$, since taking $\lambda \to 0$ will cause $\lambda \, v(-\lambda) \to 0$ (by bounded convergence) while $\eta \, v(-\lambda) \to \infty$ in the denominator. This occurs naturally when $N > P$, and incidentally would also occur if duplicate rows of $\bX$ were possible even if $P \gg N$; this latter observation makes intuitive sense, as we could then identify $\gamma_0$ using exact-matching on the $X_i$'s. 


When $P > N$ the only hope for non-negligible bias is for the eigenvalues of $\uS$ to be heavily concentrated near $0$. As $\uS$ has the same non-zero eigenvalues as the sample covariance $S = \bX^\top\bX/N$ this means we should hope for strong colinearities among the covariates. A particularly unfavorable setting is $\Sigma = \Identity$, where the Marchenko-Pastur theorem \citep[see, e.g.,][Theorem 2.13]{couillet2011random} states that if $r \ge 1$ then $F(dx)$ has density $q(\lambda) = \frac{\sqrt{(b - \lambda)(\lambda-a)}}{2 \, \pi \, \lambda} I(a < \lambda < b)$ where $(a,b) = (1 \pm \sqrt{r^{-1}})^2$; this places the bulk of the eigenvalues rather far from $0$. In Section~\ref{sec:when} we show that much better results are obtained when the $X_i$'s follow a latent factor model.

\subsection{Selection Bias Dogmatism for Semiparametric Regression}
\label{sec:application-to-semiparametric-regression}

\begin{figure}
  \centering
  \includegraphics[width=.9\textwidth]{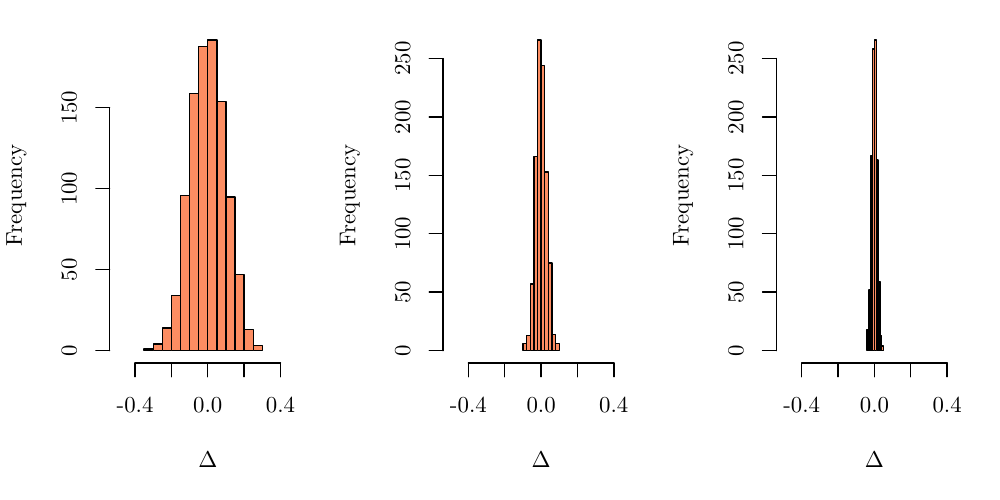}
  \caption{Prior distribution of $\Delta$ for the BART model in Section~\ref{sec:application-to-semiparametric-regression} for $P \in \{1,10,50\}$.\label{fig:Concentration}}
\end{figure}

In the semiparametric regression problem with missing data the selection bias parameter is given by 
\begin{math}
    \label{eq:semiparametric-bias}
    \Delta
    =
    \frac{\Cov_\theta\{\beta(X_i), \phi(X_i)\}}{\E_\theta\{\phi(X_i)\}}.
\end{math}
Figure~\ref{fig:Concentration} gives a sense of what to expect for nonparametric priors. In this figure, $\beta(x)$ and $\Phi^{-1}\{\phi(x)\}$ are given independent BART priors \citep{chipman2010bart, hill2011bayesian} where $\Phi(\cdot)$ is a probit function. We see that as $P$ increases the variance of $\Delta$ decreases substantially. As in the setting of ridge regression, this is troubling both because (i) it will typically violate our prior beliefs about $\Delta$ for large $P$ and (ii) given the definition of $\Delta$ as $\E_\theta(Y_i \mid A_i = 1) - \E_\theta(Y_i)$ there is no reason for our prior beliefs to be dependent on the number of confounders we happen to have measured.

For convenience we will assume that $\phi(x)$ has a point-mass prior at some $\phi_0$ and that $\beta$ has a Gaussian process prior \citep{rasmussen2005gaussian}.
Recall that $\beta \sim \GP(m, \kappa)$ means that, for any finite collection $(x_1, \ldots, x_M)$, we have $\big(\beta(x_1), \ldots, \beta(x_M)\big)^\top \sim \Normal(\bm m, \bm K)$ where $\bm m = \big(m(x_1), \ldots, m(x_M)\big)^\top$ and $\bm K$ has $(j,k)^{\text{th}}$ entry $\kappa(x_j, x_k)$. Gaussian processes have been proposed as priors for causal inference by several authors \citep{ray2020semiparametric, ren2021bayesian} and they are particularly easy to study theoretically. 

The relevant Hilbert space for applying the orthogonality principle is $\Ell_2(F_X)$, the space of square-integrable functions $\{g: \int g^2 \ dF_X < \infty  \}$ under the usual inner product  $\langle \beta, \phi \rangle = \int \beta(x) \, \phi(x) \ F_X(dx)$, with $F_X$ denoting the distribution of $X_i$. Let $\bar \beta(x) = \beta(x) - \int \beta(x) \ F_X(dx)$ and $\bar \phi(x) = \phi(x) - \int \phi(x) \ F_X(dx)$, and define the normalizations of these functions by $\widetilde \beta(x) =  \bar\beta(x) /  \|\bar\beta\|$ and $\widetilde \phi(x) = \bar\phi(x) / \|\bar\phi\|$. The following proposition shows that the selection bias is controlled by $\langle \widetilde\beta,\widetilde\phi \rangle$, implying that the orthogonality principle is in effect.
\begin{proposition}
  \label{prop:op-angle}
  Suppose $\beta \sim \GP(m, \kappa)$ such that $\sup_P \E\{\beta(X_i)^2\} < \infty$ and that there exists $\delta > 0$ such that $\E\{\phi(X_i)\} \ge \delta$ as $P \to \infty$. Then
  \begin{math}
    \Delta 
    =
    \frac{\|\bar\beta\| \, \|\bar\phi\|}{\E_\theta\{\phi(X_i)\}}
    \ 
    \langle \widetilde \beta, \widetilde \phi\rangle
     =
     O_p(\langle \widetilde\beta,\widetilde\phi \rangle).
  \end{math}
\end{proposition}


The question now is how quickly $\langle \widetilde\beta,\widetilde\phi\rangle$ tends to $0$. In the nonparametric case, in addition to the dimension ($P$) and distribution of the covariates ($F_X$), this will also depend on the \emph{smoothness} of $\widetilde\beta$ dictated by the covariance function $\kappa(\cdot,\cdot)$.
Note that $\bar\beta = \beta - \int \beta \ dF_X$ is also a Gaussian process with covariance function 
\begin{align*}
    \bar \kappa(x,x')
    =
    \kappa(x,x') - \int \kappa(x, z) \ F_X(dz) - \int \kappa(x', z) \ F_X(dz) + \iint \kappa(x,x') \ F_X(dx) \ F_X(dx').
\end{align*}
The following proposition explicitly calculates the prior distribution of $\Delta$.

\begin{proposition}
  \label{prop:gp-kl}
  Let $\beta \sim \GP\{0, \tau^2_\beta \, \rho(\cdot,\cdot)\}$ where $\rho(\cdot,\cdot)$ is a correlation function. Then $\Delta \sim \Normal(0, c)$ where $c$ is
  \begin{align*}
      \frac{\tau^2_\beta}{\E\{\phi(X_i)\}^2} \iint \bar\phi(x) \, \bar\phi(x') \, \bar \rho(x,x') \ F_X(dx) \ F_X(dx')
    =
      \frac{\tau^2_\beta}{\E\{\phi(X_i)\}^2}
          \sum_{j=1}^\infty \lambda_j \Cov\{\phi(X_i), v_j(X_i)\}^2
  \end{align*}
  and $\rho(x, x') = \sum_{j=1}^\infty \lambda_j \, v_j(x) \, v_j(x')$ is the Karhunen–Loève expansion of $\rho(x,x')$ in $\Ell_2(F_X)$.
\end{proposition}
We see that the only way for $c$ to be non-negligible is for $\phi$ to be highly correlated with some of the leading eigenfunctions of $\rho$; this suggests, at a minimum, that the kernel $\rho(x,x')$ should not be chosen in a manner which does not reference $\phi(\cdot)$. The next proposition shows that, should we not incorporate $\phi$ into $\rho(x,x')$, the resulting kernel can be universally poor: the selection bias can decay exponentially in $P$ irrespective of $\phi$.

\begin{proposition}
  \label{prop:gp-con}
  Consider the setup of Proposition~\ref{prop:gp-kl} where the covariance function $\rho(x,x')$ is given by the Gaussian kernel $\rho(x,x') = \exp\{-(x-x')^\top H^{-1} (x - x') / 2\}$ for some bandwidth matrix $H$ and suppose (i) $X_i \iid \Normal(0, \Sigma)$, (ii) $\det(\Sigma)^{1/P}/\det(H)^{1/P}$ is bounded away from $0$, and (iii) there exists a $\delta > 0$ such that $\E\{\phi(X_i)\} \ge \delta$ for all $P$. Then $\Delta \sim \Normal(0, c)$ where $c \le \exp\{-CP\}$ for some constant $C > 0$ which is independent of $\phi$. In particular, this occurs if either:
  \begin{enumerate}
  \item[(a)] $H = k \, \Sigma$ for some $k > 0$ with $k^{1/P}$ bounded and $\Sigma$ full-rank; or
  \item[(b)] $H = \xi \, \Identity$ and $\xi^{-1} \prod_j \lambda_j^{1/P}$ is bounded away from $0$.
  \end{enumerate}
\end{proposition}

Proposition~\ref{prop:gp-con} shows that the Gaussian kernel is dogmatic in a \emph{uniform} sense: no matter how favorably $\phi(x)$ is selected, the Gaussian kernel makes the prior variance on $\Delta$ decrease exponentially in $P$. Moreover, this exponential decay holds for some potentially-desirable choices of the bandwidth matrix $H$ (both when $H$ is aligned with $\Sigma$ and when the kernel is isotropic).

\subsection{Selection Bias Dogmatism for Spike-and-Slab Priors}

A common strategy for dealing with the $N \ll P$ setting in linear regression is to use a sparsity inducing spike-and-slab prior like the one described in Section~\ref{sec:three}. Even when we impose sparsity, however, serious problems occur for the selection bias prior. Suppose that $\Sigma = \sigma^2_x \Identity$ and let $\dfrak_j^\beta = I(\beta_j \ne 0)$ and $\dfrak^\phi_j = I(\phi_j \ne 0)$. Then we can write the selection bias as
\begin{math}
    \Delta(a) = a
    \frac{\sum_{j: \mathfrak{d}^\beta_j = \mathfrak{d}^\phi_j = 1} \sigma^2_x \, \phi_j \, \beta_j}
         {\sigma^2_a + \sum_{j: \mathfrak{d}^\phi_j = 1} \sigma^2_x \, \phi_j^2}.
\end{math}
In this case, the denominator (by the law of large numbers) will be of order $\sum_j \dfrak_j^\phi \equiv D_\phi$ while the numerator (by the central limit theorem) will be of order $D_{\phi \cap \beta}^{1/2}$ where $\sum_j \dfrak_j^\phi \, \dfrak_j^\beta \equiv D_{\phi \cap \beta}$. If we now use independent spike-and-slab priors for $\beta$ and $\phi$ which are calibrated to have on average $Q$ variables, we expect $D_\phi \approx Q$ while $D_{\phi \cap \beta} \approx Q/P$, so that the selection bias will be a-priori negligible in high-dimensional sparse settings in which spike-and-slab priors are applied. Hence, even if sparsity is expected (but \ref{ig3} is otherwise in effect) we run into essentially the same problem as with ridge regression: the prior on $\Delta(a)$ regularizes it towards zero.

\section{Specifying Priors Which Violate \ref{ig3}: Z-Priors}
\label{sec:specifying}

Part of the appeal of \ref{ig3} is that the treatment $A_i$ plays no role in posterior sampling of the parameter of interest $\beta$. This is computationally convenient because the updates for $\beta$ and $\phi$ in (say) a Markov chain Monte Carlo experiment can be carried out independently. It also prevents the phenomenon of \emph{model feedback} from occurring, wherein misspecification of the $A_i$-model can result in inconsistent estimation of $\beta$ even when the $Y_i$-model is correctly specified \citep{robins1997toward, zigler2013model}.

\begin{figure}
    \centering
    \includegraphics[width=.7\textwidth]{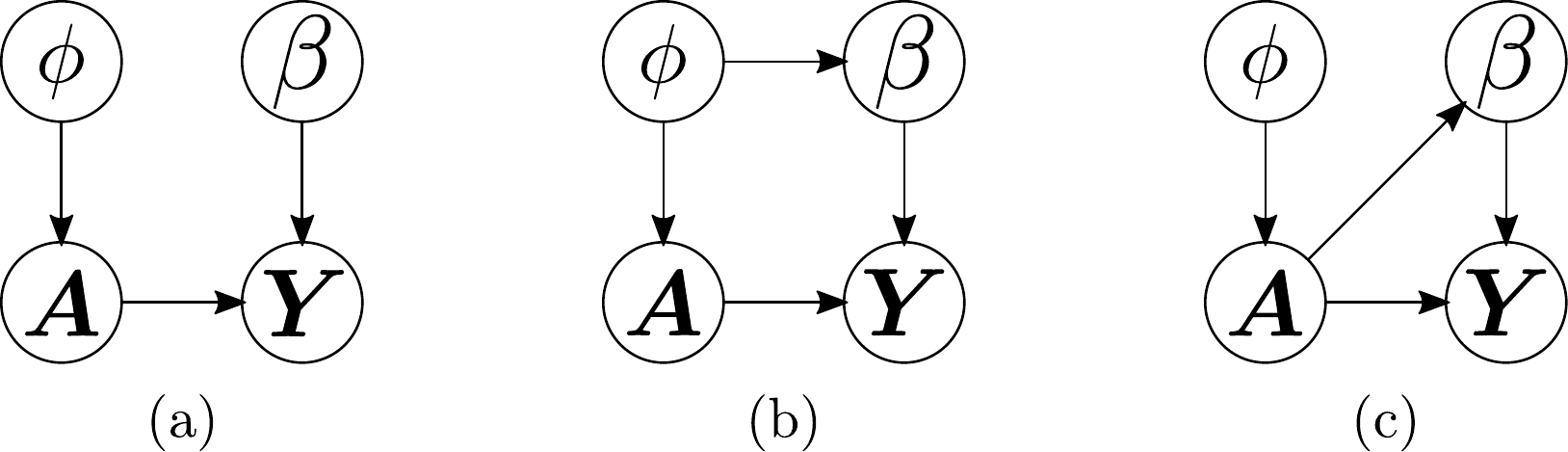}
    \caption{Directed acyclic graphs showing different conditional independence structures for model and prior specification; (a) shows the graph implied by \ref{ig3}, (b) shows the graph implied by \eqref{eq:obvious}, and (c) shows the graph implied by \eqref{eq:quasi}.}
    \label{fig:factorizations}
\end{figure}

Arguably the most natural way to specify a Bayesian model violating \ref{ig3} is to use the factorization
\begin{align}
    \label{eq:obvious}
    \pi(\phi) \, \pi(\beta \mid \phi) \, f_\phi(\bA \mid \bX) \, f_\beta(\bY \mid \bA, \bX).
\end{align}
That is, we specify the model in the usual way, but make $\beta$ dependent on $\phi$. An alternative approach, which is less natural but far more convenient, is to use the model specification
\begin{align}
    \label{eq:quasi}
    \pi(\phi, \beta, \bA, \bY \mid \bX)
    =
    \pi(\phi) \, f_\phi(\bA \mid \bX) \, \pi(\beta \mid \bA, \bX) \, f_\beta(\bY \mid \bA, \bX).
\end{align}
This approach is argued for by \citet{hahn2020bayesian} who refer to specifications like \eqref{eq:quasi} as \emph{Zellner priors} due to the fact that the prior for $\beta$ is allowed to depend on the design matrix $(\bA, \bX)$ of $\bY$ (as is the case for Zellner's famous $g$-prior). In order to avoid confusion with Zellner's $g$-prior, we refer to priors of this form as Z-priors. Figure~\ref{fig:factorizations} shows a schematic comparing these two factorizations with \ref{ig3}. This prior makes $\beta$ and $\phi$ \emph{conditionally independent} given $(\bA, \bX)$ so that it retains the principle advantage of \ref{ig3}: feedback between the $Y_i$ and $A_i$ models is severed, and the updates for $(\beta, \phi)$ are no longer coupled. Note that \eqref{eq:quasi} induces a dependent prior of the form
\begin{math}
    \pi(\phi, \beta) = 
    \pi(\phi) \, \int \pi(\beta \mid \bA, \bX) \, f_\phi(\bA \mid \bX) \ d\bA \ dF_X,
\end{math}
so that \ref{ig3} is violated.

To illustrate the point, in Section~\ref{sec:direct} we will consider a prior for the ridge regression problem which is of the form $\beta \sim \Normal(\omega \, \phi, \tau^2_\beta \, \Identity)$ where $\omega$ is given a diffuse prior. This prior conforms to \eqref{eq:obvious}. In our actual experiments, however, we use the prior $\beta \sim \Normal(\omega \, \widehat \phi, \tau^2_\beta \, \Identity)$ where $\widehat \phi = \E(\phi \mid \bA, \bX)$ is a data-adaptive ridge estimator. This prior is of the form \eqref{eq:quasi} and is trivial to implement in a two-stage fashion: fit the model for $A_i$, compute $\widehat \phi$, and plug this into the prior for $\beta$ when fitting the $Y_i$-model.

In our experience, some Bayesians feel uneasy about using priors like $\beta \sim \Normal(\omega \widehat \phi, \tau^2_\beta)$ because it ``understates the uncertainty'' in $\beta$ due to the fact that it appears to use a plug-in estimate of $\phi$ rather than $\phi$ itself. While it is true that using the actual value of $\phi$ rather than an estimate typically (although not always! see \citealp{hirano2003efficient}) results in improved Frequentist performance, the justification of this prior as being of the form \eqref{eq:quasi} shows that there is no explicit violation of the Bayesian calculus in using this prior. For example, for the ridge regression Z-prior discussed above we can explicitly derive the induced prior on $\pi(\beta, \phi , \omega \mid \bX)$ as
\begin{math}
    \pi(\beta, \phi, \omega \mid \bX) = \pi(\phi, \omega) \, 
    \Normal\big(\beta\mid \omega \, \phi , \tau^2_\beta \, \Identity + \sigma^2_a (\bX^\top \bX)^{-1}\big).
\end{math}

\section{Correcting for Dogmatism}
\label{sec:correcting-for-dogmatism}


\subsection{Direct Priors for Ridge Regression}
\label{sec:direct}

A simple approach to addressing dogmatism in the ridge regression setting is to note that we can make $\beta^\top \Sigma \phi$ large by encouraging $\beta$ to align with $\phi$. For example, we might center $\beta$ on $\phi$ by taking $\beta \sim \Normal(\omega \phi, \tau^2_\beta \, \Identity).$ 
Doing this, we now have
\begin{math}
    \Delta(a) 
    = 
    a \frac{\phi^\top \Sigma b}{\sigma^2_a + \phi^\top \Sigma \phi}
    +
    a\, \omega \frac{\phi^\top \Sigma \phi}{\sigma^2_a + \phi^\top \Sigma \phi},
\end{math}
where $b \sim \Normal(0, \tau^2_\beta \, \Identity)$. By the same argument as in Proposition~\ref{prop:ridge-clt}, the first term is $O_p(P^{-1/2})$; the second term, however, does not tend to $0$ as $P \to \infty$, preventing prior dogmatism from taking hold. By centering the prior for $\beta$, we can now specify a \emph{direct} prior on $\Delta(a)$ by placing a prior on $\omega$. For example, we can express prior ignorance about the degree of selection bias by placing a flat prior on $\omega$.

This approach is related to the targeted maximum likelihood estimation strategy of introducing a ``clever covariate''  into the outcome model to account for selection \citep[see, e.g.,][Section 4.2.1]{van2011targeted}. The parameterization $\beta = b + \omega \phi$ gives
\begin{math}
  Y_i(a) = \beta_0 + X_i^\top b + \omega (X_i^\top \phi) + \gamma \, a + \epsilon_i(a),
\end{math}
which effectively introduces the new covariate $\widehat A_i = X_i^\top\phi$ into the model. A related idea proposed by \citep{hahn2018regularization} is to replace $a$ in the outcome model with the residual $(a - \widehat A_i)$, which is equivalent to setting $\omega = -\gamma$.

\subsubsection*{Bias of the Z-prior Estimate under HDA and REM}

In practice, we will not usually use the direct prior described above; instead, we will use a Z-prior which plugs in a point-estimate of the clever covariate $\widehat A_i = X_i^\top\widehat\phi$. Assuming HDA and REM, the bias of the Bayes estimator under the Z-prior can be shown to be
\begin{align*}
    \E(\widetilde\gamma - \gamma_0)
    =
  \E\left\{\frac{\bA^\top (\Identity - \widehat \Psi \bar H \widehat \Psi^\top)\Psi \binom{\omega_0}{b}}{\bA^\top (\Identity - \widehat \Psi \bar H \widehat \Psi^\top)\bA}\right\}
\end{align*}
where $\widehat \Psi = [\bX\widehat\phi, \bX]$, $\Psi = [\bX\phi,\bX]$ and $\bar H^{-1} = \widehat \Psi^\top\widehat \Psi + N\lambda\left(\begin{smallmatrix}0 & 0 \\ 0 & \Identity \end{smallmatrix}\right)$. If one happened to know the exact value of $\phi$ and set $\widehat \phi = \phi$ then it is easy to show that the resulting Bayes estimator $\widetilde\gamma$ is unbiased for $\gamma$, irrespective of the prior for $b$. It is also easy to show that if we take $\widehat\phi=\widetilde\phi$ where $\widetilde\phi$ is the Bayes estimate under the correct prior $\widetilde \phi = \E(\phi \mid \bA, \bX)$ then $\widetilde\gamma$ remains unbiased.

The selling point now is that there are moderate values of $\lambda$ for which ridge regression will have $0$ bias, a situation which was not possible with the naive prior. In practice, under REM we will know neither $\phi$ nor $\widetilde\phi$ (because we will not know the signal level $\tau_\phi^2$). Data is typically quite informative about $\tau^2_\phi$, however, and we have had success placing a prior on $\tau^2_\phi$ to obtain nearly unbiased estimates. This is seen in Figure~\ref{fig:bias_compare} ($\Sigma = \Identity$), where the point on the solid line corresponds to the bias if we plug in a Bayes estimate of $\tau^2_\phi$ to construct a ridge estimator. 

It is also possible to show that the asymptotic bias of the Z-prior when $\widehat \phi$ is estimated with ridge regression is
\begin{align*}
  \left(
  \omega_0 \, \lambda 
  \left[
  \frac{\psi_{11}}{\eta} - \frac{\psi_{22}}{\eta} \frac{(\psi_{21} + \psi_{22}/\eta)}{(\psi_{32} + \psi_{33}/\eta)}
  \right]
  \right) / 
  \left(
  {\frac{1-r}{r} + \lambda 
  \left[
  \psi_{10}
  + \frac{\psi_{11}}{\eta}
  - \frac{(\psi_{21} + \psi_{22}/\eta)^2}
  {(\psi_{32} + \psi_{33}/\eta)}
  \right]
  }
  \right)
\end{align*}
where $\psi_{jk} = \int_0^\infty \frac{x^k}{(x + \lambda)^j} \ G(dx)$ and $G(dx)$ is the empirical spectral distribution corresponding to $\bX^\top\bX/N$ (also known as the companion spectral distribution to $F(dx)$). Each of the $\psi_{jk}$'s can be computed by noting the recursive identity $\psi_{jk} = \psi_{j-1,k-1} - \lambda \psi_{j,k-1}$ and the fact that $\psi_{j0} = m^{(j-1)}(-\lambda) / (j-1)!$ where $m(z)$ is the Stieltjes transform of $G(dx)$.

A plot of the bias is given in Figure~\ref{fig:bias_compare} when $\Sigma = \Identity$, and we numerically our bias formula in the Supplementary Material.
Rather curiously, for $r > 1$ we see that the bias of the Z-prior estimate is ill-behaved near $0$; fortunately, we can estimate $\tau^2$ accurately enough to avoid this region.

\subsubsection*{Evaluation of the Direct Z-Prior via Simulation}

In order to determine if there is any benefit to using the Z-prior with the prior on $\omega$ being $\Uniform(-\infty, \infty)$ relative to either (i) the naive ridge regression prior or (ii) the approach of \citep{hahn2018regularization} which we call the ``debiased'' approach (equivalent to fixing $\omega = -\gamma$), we conducted a simulation study. In all cases we set $N = 200$ and $P = 1000$ so that $N \ll P$. We consider a dense model with $\phi = (1,\ldots,1) / \sqrt P$, a randomly chosen $\beta$ vector $\beta_j \iid \Normal(0, P^{-1})$, and $\sigma^2_a = \sigma^2_y = 1$. The methods differ in the treatment effect size $\gamma$ and the degree to which the coefficients are shifted in the direction of $\phi$. We considered three simulation settings.

\begin{description}[style=unboxed,leftmargin=0cm]
  \item[Random] Both $\gamma$ and $\omega$ are $\Normal(0,1)$ random variables and differ for each replication of the experiment.
  \item[Fixed] We set $\gamma = 2$ and $\omega = -\gamma / 4$ so that $\beta$ is shifted in the direction of $\phi$, but not by the amount implied by the debiased approach.
  \item[Debiased] We set $\gamma = 2$ and $\omega = -2$ so that $\beta$ is shifted in the direction of $\phi$ by exactly the amount implied by the debiased approach.
  \item[Naive] We set $\gamma = 2$ and $\omega = 0$ so that the model corresponds precisely to the naive ridge model.
\end{description}

The simulation was replicated 200 times for each setting. We evaluated each procedure according to the following criteria. \textbf{Coverage:} The proportion of nominal 95\% credible intervals which capture the true value of $\gamma$. \textbf{Width:} The average width of the nominal 95\% credible interval. \textbf{Avg SE:} The average estimated standard error from the model, i.e., the the posterior standard deviation of $\gamma$ averaged over all replications. \textbf{RMSE:} The root mean squared error in estimating $\gamma$ with the Bayes estimator $\widehat\gamma$.

\begin{table}
\centering %
\begin{tabular}{cccccc}
\toprule 
Setting                            & Method   & Coverage & Width         & Avg SE & RMSE \\
\midrule 
\multirow{3}{*}{$\omega\sim\Normal(0,1)$}            & Direct   & 0.94     & {0.39} & 0.10   & 0.10 \\
                                   & Debiased & 0.94     & 0.51          & 0.13   & 0.12 \\
                                   & Naive    & 0.19     & 0.31          & 0.08   & 0.49 \\
\cmidrule{2-6} \cmidrule{3-6} \cmidrule{4-6} \cmidrule{5-6} \cmidrule{6-6} 
\multirow{3}{*}{$\omega=\gamma/4$} & Direct   & 0.95     & {0.39} & 0.10   & 0.11 \\
                                   & Debiased & 0.94     & 0.54          & 0.13   & 0.14 \\
                                   & Naive    & 0.12     & 0.29          & 0.07   & 0.25 \\
\cmidrule{2-6} \cmidrule{3-6} \cmidrule{4-6} \cmidrule{5-6} \cmidrule{6-6} 
\multirow{3}{*}{$\omega=-\gamma$}   & Direct   & 0.96     & {0.39} & 0.10   & 0.11 \\
                                   & Debiased & 0.95     & 0.39          & 0.10   & 0.11 \\
                                   & Naive    & 0.00     & 0.40          & 0.10   & 0.96 \\
\cmidrule{2-6} \cmidrule{3-6} \cmidrule{4-6} \cmidrule{5-6} \cmidrule{6-6} 
\multirow{3}{*}{Naive ($\omega=0$)}             & Direct   & 0.95     & {0.39} & 0.10   & 0.11 \\
                                   & Debiased & 0.93     & 0.63          & 0.16   & 0.16 \\
                                   & Naive    & 0.90     & 0.28          & 0.07   & 0.08 \\
\bottomrule
\end{tabular}
\caption{Comparison of different approaches for estimating $\gamma$ under
differing levels of selection bias. Direct denotes the approach which
sets $\omega\sim\protect\operatorname{Flat}$, Debiased denotes the approach of
\cite{hahn2018regularization}.\label{tab:hahn}}
\end{table}

Results are compiled in Table~\ref{tab:hahn}. The direct and debiased approaches always attain the nominal coverage level, while the naive approach does not come close when the selection bias is non-negligible. We also see that the debiased approach will generally require substantially larger intervals than the direct approach to cover at the appropriate rate; the only exception is when $\omega = -\gamma$, which is to be expected as this setting agrees exactly with the debiased prior. When the naive ridge model actually holds (i.e., $\omega = 0$) we see that the naive ridge model unsurprisingly performs substantially better, and is the best in terms of RMSE, with the direct prior still outperforming the debiased prior. 


\subsection{Variable Sharing for Spike-and-Slab Priors}
\label{sec:variable-sharing}

For the variable selection prior it remains valid to include the ``clever covariate'' $X_i^\top \widehat \phi$ in the model to correct for dogmatism. However, there are other approaches one can take which make specific use of the variable selection aspect of the model. One possibility is to use \emph{shared variable selection} for the two models; in particular, we want to ensure any variable appearing in the selection model should also appear in the outcome model. To implement this, we might set $\phi_j \iid (1 - p_\phi) \, \delta_0 + p_\phi \, \Normal(0, \tau^2_\phi)$ and conditionally set $\beta_j \indep \{1 - p_\beta(\phi_j)\} \, \delta_0 + p_\beta(\phi_j) \, \Normal(0, \tau^2_\beta)$. Setting $p_\beta(\phi_j) = 1$ if $\phi_j \ne 0$ guarantees that $\beta_j$ will be included whenever $\phi_j$ is included.

We conduct a small simulation experiment to justify our claim that shared variable selection is an effective strategy for combating dogmatism. For our ground truth we consider $N = P = 200$, $\sigma^2_a = \sigma^2_y = 1$, and $\gamma = 1$. We then sample $\phi_j$ from the spike-and-slab prior with $\tau^2_\phi = 1$ and $p_\phi = 5 / 200$. We then consider four different schemes for sampling $\beta$. \textbf{Naive:} we sample $\beta_j$ from the spike-and-slab prior with $p_\beta = 5 / 200$ and $\tau^2_\beta = 1$; \textbf{Shared:} we sample $\beta_j$ from the spike-and-slab prior with $p_\beta = 5 / 200$ if $\phi_j = 0$ and $p_\beta = 1$ if $\phi_j \ne 0$; \textbf{Direct:} we sample $\beta_j$ according to the Naive prior and then add $-\phi_j$ to it; and \textbf{Both:} we sample $\beta_j$ according to the Shared prior then we add $-\phi_j$ to it.

We compare the following prior specifications.
\begin{description}[style=unboxed,leftmargin=0cm]
  \item[Naive] $\beta_j$ has a spike-and-slab prior with $p_{\beta} \equiv 5/200$ and which is independent of $\phi_j$.
  \item[Shared] We use a spike-and-slab Z-prior with $p_{\beta,j}  = 5/200$ if $\Pr(\phi_j = 0 \mid \bA, \bX) < 0.5$, and $p_{\beta,j} = 1$ otherwise.
  \item[Direct] We use the Naive prior with the additional covariate $\widehat A_i = X_i^\top \widehat \phi$ where $\widehat \phi = \E(\phi \mid \bA)$. The variable $\widehat A_i$ is included with probability $1$.
\end{description}

\begin{table}
\centering %
\begin{tabular}{llrrrrr}
\toprule 
Setting  & Method & Coverage & Width & Std. Err & RMSE & Bias\tabularnewline
\midrule
\multirow{3}{*}{Shared} & Direct  & 0.95  & 0.29  & 0.07  & 0.07  & 0.00 \tabularnewline
 & Naive  & 0.65  & 0.25  & 0.07  & 0.13  & -0.00 \tabularnewline
 & Shared  & 0.94  & 0.28  & 0.07  & 0.07  & 0.00 \tabularnewline
\cmidrule{2-7} \cmidrule{3-7} \cmidrule{4-7} \cmidrule{5-7} \cmidrule{6-7} \cmidrule{7-7} 
\multirow{3}{*}{Direct} & Direct  & 0.95  & 0.29  & 0.07  & 0.08  & -0.03 \tabularnewline
 & Naive  & 0.71  & 0.30  & 0.08  & 0.34  & -0.17 \tabularnewline
 & Shared  & 0.93  & 0.29  & 0.07  & 0.08  & -0.03 \tabularnewline
\cmidrule{2-7} \cmidrule{3-7} \cmidrule{4-7} \cmidrule{5-7} \cmidrule{6-7} \cmidrule{7-7} 
\multirow{3}{*}{Naive} & Direct  & 0.96  & 0.28  & 0.07  & 0.07  & -0.00 \tabularnewline
 & Naive  & 0.94  & 0.14  & 0.04  & 0.04  & -0.00 \tabularnewline
 & Shared  & 0.97  & 0.28  & 0.07  & 0.07  & -0.00 \tabularnewline
\cmidrule{2-7} \cmidrule{3-7} \cmidrule{4-7} \cmidrule{5-7} \cmidrule{6-7} \cmidrule{7-7} 
\multirow{3}{*}{Both} & Direct  & 0.94  & 0.29  & 0.07  & 0.07  & 0.00 \tabularnewline
 & Naive  & 0.77  & 0.27  & 0.07  & 0.12  & -0.01 \tabularnewline
 & Shared  & 0.94  & 0.28  & 0.07  & 0.07  & 0.00 \tabularnewline
\bottomrule
\end{tabular}
\caption{Results for the simulation experiment described in Section~\ref{sec:variable-sharing}}
\end{table}

Results for this simulation are given in Table~\ref{sec:variable-sharing}. Each simulation setting was replicated $200$ times. We see that the Direct and Shared methods perform essentially the same despite correcting for dogmatism in different ways --- both methods have virtually identical coverage, root-mean-squared error, interval lengths, and bias. The Naive approach, which just applies the spike-and-slab prior for $\beta$ under \ref{ig3}, performs extremely poorly by contrast, unless the data was generated under the Naive prior.

\subsection{Semiparametric Regression with Clever Covariates}
\label{sec:semiparametric-regression-with-clever-covariates}

Mimicking our strategy in Section~\ref{sec:direct} we set $\beta(x) = \beta^\star(x) + g\{\phi(x)\}$ for some choice of $g(\cdot)$, with $\beta^\star(x)$ given (say) a Gaussian process prior independent of $\phi(\cdot)$. The selection bias for this model is given by
\begin{align*}
    \Delta 
    = \frac{\Cov_\theta\{\beta^\star(X_i), \phi(X_i)\}}{\E_\theta\{\phi(X_i)\}}
    + \frac{\Cov_\theta[g\{\phi(X_i)\}, \phi(X_i)]}{\E_\theta\{\phi(X_i)\}}
    \approx
    \frac{\Cov_\theta[g\{\phi(X_i)\}, \phi(X_i)]}{\E_\theta\{\phi(X_i)\}}
\end{align*}
by the orthogonality principle. Even if we model $g(\cdot)$ nonparametrically using (say) a Gaussian process the orthogonality principle will not kick in because $g(\phi)$ is a function on $\Reals$ rather than $\Reals^P$.

There are many different choices one can make for the function $g(\cdot)$. If we are concerned strictly with obtaining good Frequentist properties, an appropriate choice is to take $g(\phi) = \phi^{-1}$; when $\phi$ is known, this guarantees $\sqrt n$-consistency. 
Alternatively, we can set $g \sim \GP(0, \kappa_g)$ with the covariance function $\kappa(\phi, \phi') = \tau^2_g \exp\{-(\phi - \phi')^2 / (2s^2_g)\}$. This choice of covariance function was noted by \citet{ren2021bayesian} to induce matching on the propensity score: individuals with similar propensity scores have their values of $g(\phi)$ shrunk together. The penalized-spline-of-propensity approach of \citet{zhou2019penalized} is similar, except that $g(\phi)$ is chosen to be a spline instead of a Gaussian process. Another approach to incorporating $\phi(x)$ is to plug it in as a regular covariate, i.e., we replace $\beta(x)$ with $\beta\{x, \phi(x)\}$. 


\subsubsection*{Simulation Experiment}

We consider the simulation setting of \citet[Section 6.1]{hahn2020bayesian} to evaluate several different approaches to correcting a Gaussian process prior for dogmatism using the usual evaluation criteria (RMSE, interval width, and bias, and coverage). We consider the model
\begin{math}
    Y_i(a) = \mu(X_i) + a \, \tau(X_i) + \epsilon_i,
    \epsilon_i \sim \Normal(0,1)
\end{math}
with the $X_{ij}$'s being iid $\Normal(0,1)$ random variables with the exception of $X_{i2}$ which is $\Bernoulli(1/2)$ and $X_{i4}$ which is uniform on $\{1,2,3\}$. We let $\mu(x)$ and $\tau(x)$ be given
\begin{align}
    \label{eq:semipar-set}
    \tau(x) =
    \begin{cases}
      3 \quad & \text{homogeneous}, \\[-1em]
      1 + 2 \, x_2 \, x_5 & \text{heterogeneous},
    \end{cases}
    \quad \text{and} \quad 
        \mu(x) =
    \begin{cases}
      1 + g(x_4) + x_1 \, x_3 \quad & \text{linear}, \\[-1em]
      -6 + g(x_4) + 6 \, |x_3 - 1| & \text{nonlinear},
    \end{cases}
\end{align}
where $g(1) = 2, g(2) = -1$, and  $g(3) = -4$. We then set $A_i \sim \Bernoulli\{\phi(X_i)\}$ with $\phi(x) = 0.8 \, \Phi\{3 \, \mu(x) / s - 0.5 \, x_1\} + 0.1$, where $s$ is the empirical standard deviation of the $\mu(X_i)$'s. In total, we consider 16 possible simulation settings, corresponding to a factorial design with $N \in \{250, 500\}$, $P\in \{5,20\}$, with four combinations of linear/nonlinear and homogeneous/heterogeneous.

We consider modeling $\E\{Y_i(a) \mid X_i = x\} = \beta(a, x)$ using a Gaussian process $\beta \sim \GP(0, \kappa)$ where the kernel function $\kappa\big((a,x), (a',x')\big)$ is given by the following choices.
\begin{description}[style=unboxed,leftmargin=0cm]
  \item[Naive] A kernel which makes no correction for the propensity score: $\kappa\big((a,x), (a',x')\big) = 100 (1 + a \, a') + \lambda \, \exp\{-b \|(a,x) - (a',x')\|^2_2\}$.
  \item[IPW-GP] A kernel which incorporates the inverse propensity score as a ``clever covariate'' which enters the model linearly: $\kappa\big((a,x), (a',x')\big) = 100 (1 + a \, a' + w \, w' + z \, z') + \lambda \, \exp\{-b \|(a,x) - (a',x')\|^2_2\}$ where $w = a / \phi(x)$ and $z = (1 - a) / (1 - \phi(x))$.
  \item[Spline-of-propensity-GP] A kernel which incorporates the propensity score using a spline basis function expansion: $\kappa\big((a,x), (a',x')\big) = 100(1 + a \, a' + \sum_k \psi_k \, \psi_k') + \lambda\exp\{-b \|(a, x) - (a',x')\|^2_2\}$ where $\psi_k = \psi_k(x)$, $\psi'_k = \psi(x')$, and $\{\psi_1, \ldots, \psi_K\}$ are natural cubic spline basis functions using $10$ knots.
  \item[Spline-of-propensity] Same as Spline-of-propensity-GP but without the Gaussian kernel.
\end{description}
In order to to separate the issue of accurately estimating the propensity scores from the benefit of using them, we assume that $\phi(x)$ is known a-priori. For all methods, the kernel hyperparameters and the standard deviation $\epsilon$ are estimated via empirical Bayes, i.e., by maximizing the marginal likelihood of the data.

\begin{figure}[t]
    \centering
    \includegraphics[width=1\textwidth]{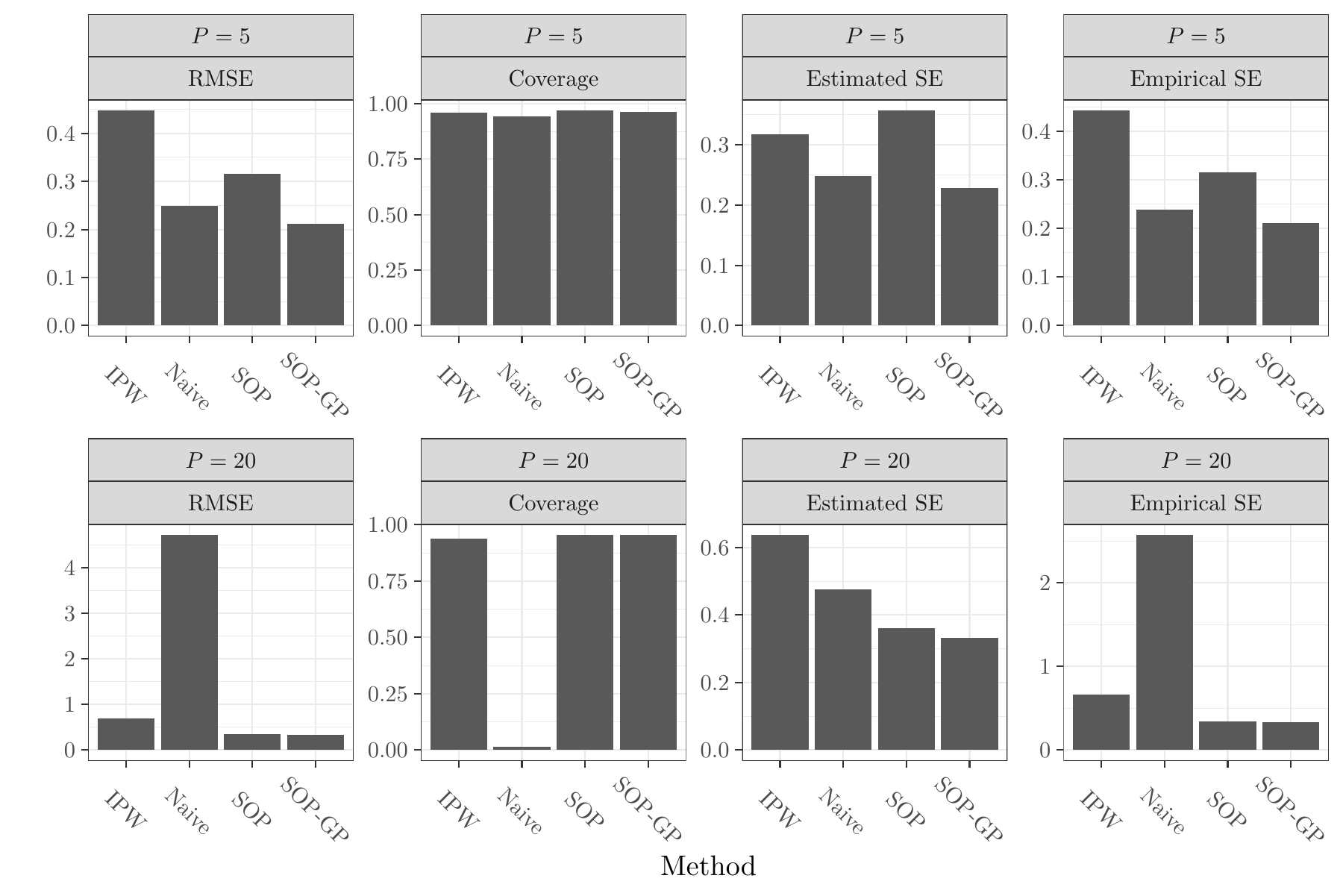}
    \caption{Results for the simulation study of Section~\ref{sec:semiparametric-regression-with-clever-covariates} in the nonlinear heterogeneous setting. Naive denotes the naive approach, IPW denotes the IPW-GP approach, SOP denotes the spline-of-propensity approach, and SOP-GP denotes the spline-of-propensity-GP approach.}
    \label{fig:np_dogma}
\end{figure}

Our main goals are to (i) determine the extent to which the Naive kernel suffers due to dogmatism, (ii) determine which of the IPW or spline approaches perform better in this case, and (iii) determine whether the propensity score alone is sufficient to produce a good estimator. A subset of the results corresponding to the nonlinear heterogeneous setting with $N = 250$ are given in Figure~\ref{fig:np_dogma}, with the remaining results deferred to the Supplementary Material. Summarizing these results, we find (i) that the Naive kernel performs well when $P = 5$ where dogmatism is mild, but breaks down completely when $P = 20$; (ii) that the IPW-GP and spline-of-propensity-GP approaches perform comparably in terms of coverage, but that the spline-of-propensity-GP generally produces smaller standard errors and RMSEs, suggesting that the spline-of-propensity approach is more stable while accomplishing the same goals as IPW methods; and (iii) that the spline-of-propensity-GP produces smaller standard errors and RMSEs than the spline-of-propensity approach, indicating that there is some benefit to going beyond simply adjusting for the propensity score.

\section{Factors Mitigating Dogmatism}
\label{sec:when}

In demonstrating the issue of dogmatism we made use of the orthogonality principle, which we noted only holds when ``other structures'' are not present. For example, we introduced additional structure into the model by making $\beta$ and $\phi$ dependent in Section~\ref{sec:correcting-for-dogmatism}, giving us direct control of the prior on $\Delta(a)$.

Another possible source of structure is dependence structure in $X_i$. The benefit of this is evident in Proposition~\ref{prop:ridge-clt} and Theorem~\ref{thm:ridge}, where the spectral distribution of $\Sigma$ figures prominently. We examine the role of dependence structure in the ridge and semiparametric regression problems.

\subsection{Dependence Structure and Ridge Regression}


To understand the role of $\Sigma$ in ridge regression, we consider a \emph{latent factor model} which takes $X_i = \Lambda \eta_i + \sigma_x \, \nu_i$ where $\Lambda \in \Reals^{P \times L}$ is a matrix of factor loadings and $\eta_i \in \Reals^L$ is an $L$-dimensional vector of latent factors for observation $i$. If $\sigma_x = 0$ in this model then $X_i$ is restricted to be in the $L$-dimensional subspace $\Span(\Lambda)$; similarly, if $\sigma_x$ is small, then $X_i$ lies very close to $\Span(\Lambda)$. 

We first examine the induced prior on the selection bias parameter for such a $\Sigma$. Assuming $\eta_i \sim \Normal(0, \Identity)$ and $\nu \sim \Normal(0, \Identity)$, we have $\Var(X_i) \equiv \Sigma = \Lambda \Lambda^\top + \sigma_x^2 \Identity$. Letting $\kappa_1, \ldots, \kappa_L$ denote the $L$ non-zero eigenvalues of $\Lambda \Lambda^\top$, Proposition~\ref{prop:ridge-bias} gives
\begin{align*}
    \Delta(a)
    &=
    a \frac{\sum_{j=1}^L (\kappa_j + \sigma^2_x) \, W_j \, Z_j}{\sigma^2_a + \sum_{j=1}^L (\kappa_j + \sigma^2_x) \, Z_j^2} + 
    a \frac{\sum_{j=L+1}^P \sigma^2_x \, W_j \, Z_j}{\sigma^2_a + \sum_{j=L+1}^P \sigma_x^2 \, Z_j^2}
  \approx
    a \frac{\sum_{j=1}^L \kappa_j \, W_j \, Z_j}{\sigma^2_a + \sum_{j=1}^L \kappa_j \, Z_j^2}
\end{align*}
where $W_j \iid \Normal(0, \tau^2_\beta)$ and $Z_j \iid \Normal(0, \tau_\phi^2)$. The approximation holds when $\sigma_x$ is near zero so that $\Sigma$ is approximately low-rank. Because the left-hand-side depends only on $L$ rather than $P$, we expect $\Delta(a)$ to be roughly of order $L^{-1/2}$ rather than $P^{-1/2}$ for the ridge regression prior. Hence, even if $P \gg N$, we may still avoid dogmatism if $L \ll N$. 

\begin{figure}[t]
    \centering
    \includegraphics[width=.9\textwidth]{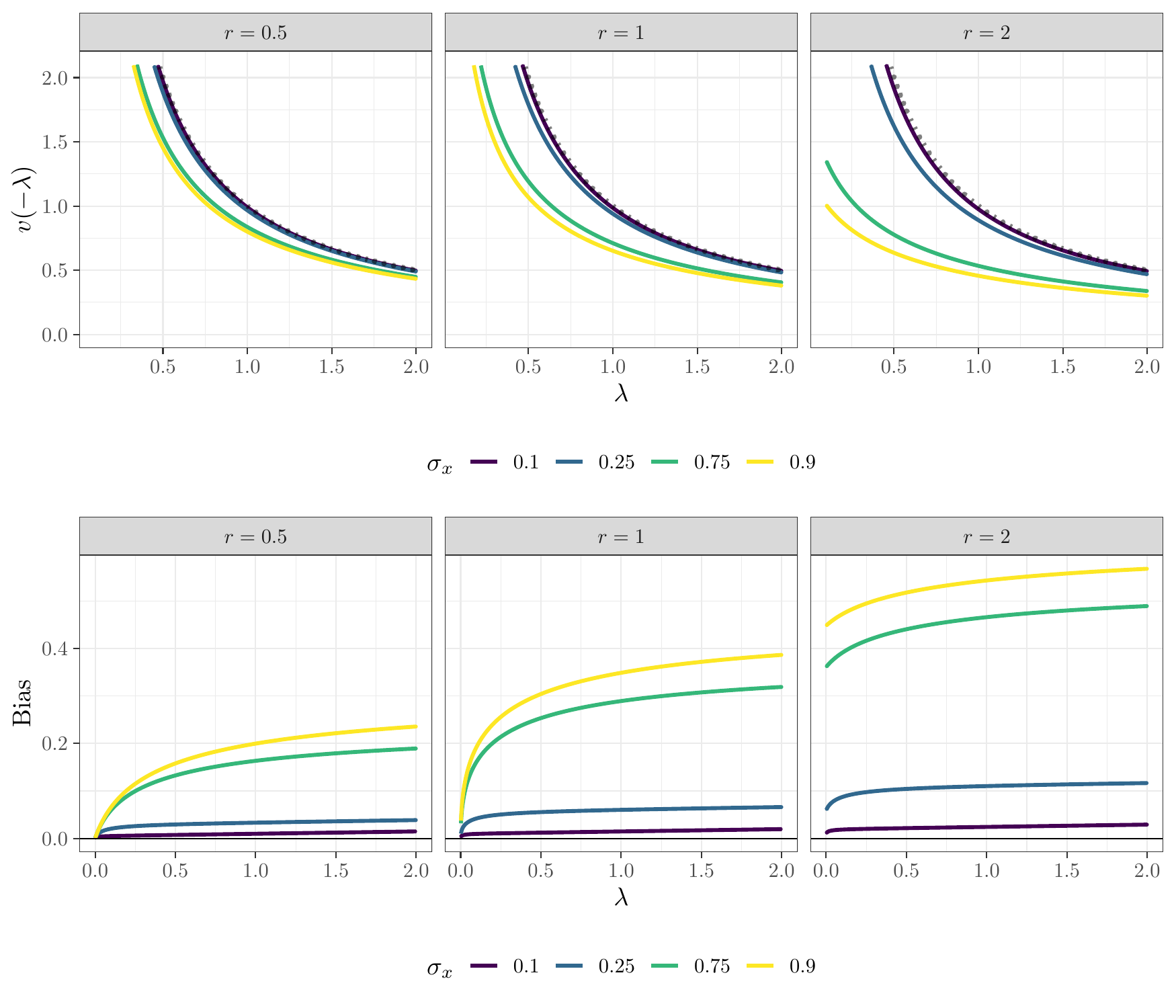}
    \caption{Top: $v(-\lambda)$ for the latent factor model for different $(r,\sigma_x)$ with a dashed line giving the ideal $v(-\lambda) = \lambda^{-1}$. Bottom: the associated bias of the ridge regression estimator.}
    \label{fig:ridge_low_rank_explore}
\end{figure}

Revisiting Theorem~\ref{thm:ridge} we can characterize the effect of $\Sigma$ on the bias. Recall that Theorem~\ref{thm:ridge} relates the bias of the naive ridge estimator to the empirical spectral distribution $F(dx)$ of $\bX\bX^\top/N$ through the Stieltjes transform $v(-\lambda) = \int_0^\infty \frac{F(dx)}{x + \lambda}$. Smaller values of $\lambda$ such that $v(-\lambda) \approx \lambda^{-1}$ result in small bias, which occurs when $F(dx)$ places substantial mass near $0$. In Figure~\ref{fig:ridge_low_rank_explore} we plot both $v(-\lambda)$ and the bias for the latent class model with $L = 5$ and the entries of $\Lambda$ being iid $\Normal(0,1)$. We see that as $\sigma_x$ decreases we have substantially less bias. The ridge regression estimator is able to leverage the fact that $X_i$ is approximately in $\Span(\Lambda)$ without this being explicitly encoded into the model. In the extreme case where $\sigma_x = 0$ this is easy to see, as we can write $\bX = \bE \bm D^{1/2} \Gamma^\top$ where $\bE \in \Reals^{N\times L}$ has iid $\Normal(0,1)$ entries, $\Gamma \in \Reals^{P \times L}$ is semi-orthogonal, $\bm D \in \Reals^{L\times L}$ is diagonal, and $\Sigma = \Gamma \bm D \Gamma^\top$. We can then rewrite $\bX \beta = \bE \bm D^{1/2} \Gamma^\top \beta = \bE \zeta$ where $\zeta \sim \Normal(0, \bm D \, \tau^2_\beta)$. Hence performing ridge regression on the $N \times P$-dimensional $\bX$ is exactly equivalent to performing ridge regression on the lower-dimensional matrix $\bE$, with $\zeta_\ell$ having variance proportional to the $\ell^{\text{th}}$ non-zero eigenvalue of $\Sigma$.

\begin{figure}
  \centering
  \includegraphics[width=.47\textwidth]{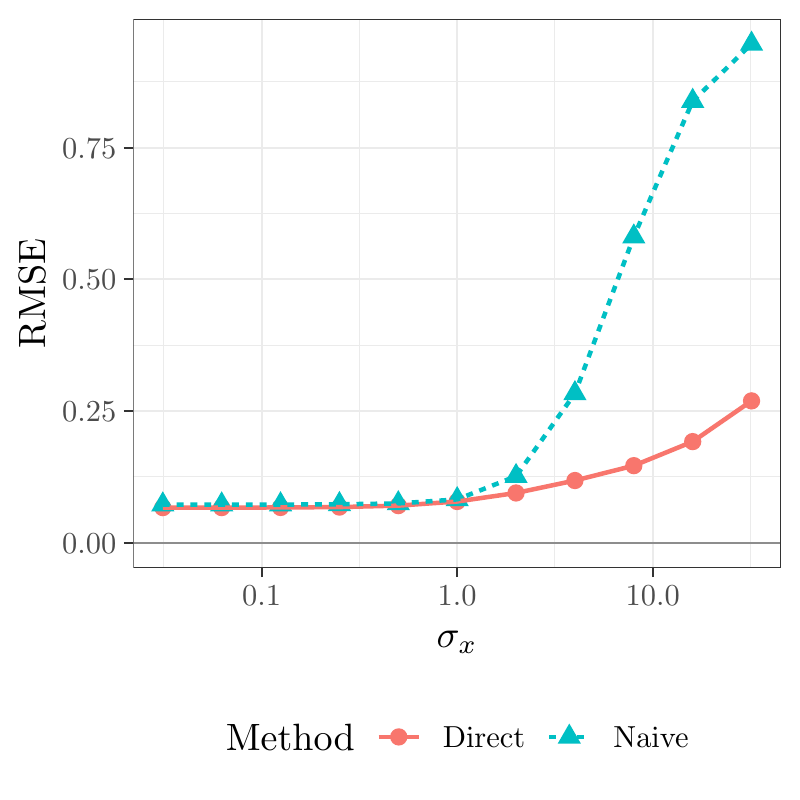}
  \caption{
    Root mean squared error in estimating $\gamma$ for direct and naive methods as a function of $\sigma_x$.
    \label{fig:manifold}}
\end{figure}

Figure~\ref{fig:manifold} shows the root mean squared error of the direct approach we proposed in Section~\ref{sec:direct} and the naive ridge regression estimator. To generate this figure, we applied the two approaches under the following conditions: $\sigma^2_a = \sigma^2_y = 1$; $L = 5$; $N = P = 200$; $\Lambda_{p\ell} \iid \Normal(0,1)$; $\gamma = 1$; and both $\beta$ and $\phi$ chosen so that $X_i^\top \beta = X_i^\top \phi = \sum_{\ell = 1}^L \E(\eta_{i\ell} \mid X_i)$.

  

\subsection{Dependence Structure and Semiparametric Regression}

  We present evidence that the results for ridge regression is part of a much more general phenomenon which also holds in nonparametric problems: namely, that low-dimensional structure can shield us from the effects of prior dogmatism. Rather than assuming that $X_i$ is near a hyperplane, we now assume that $X_i$ is concentrated near a smooth manifold of intrinsic dimension $L$ in $\Reals^P$. Specifically, we take
  \begin{math}
      \widetilde X_i = \Lambda(\eta_i) + \sigma_x \, \epsilon_i,  \epsilon_i \sim \Normal(0,1)
  \end{math}
  where $\Lambda: \Reals^L \to \Reals^P$ is nonlinear; we then scale $\widetilde X_{ij}$ by its standard deviation to get $X_{ij}$, so that $\sigma_x$ indexes how close $X_i$ is to $\sM = \{x : x = \Lambda(\eta), \eta \in \Reals\}$. We randomly generated $\Lambda(\eta) = \big(\Lambda_1(\eta), \ldots, \Lambda_P(\eta)\big)^\top$ by generating $P$ independent Gaussian processes using the kernel function $\rho(\eta, \eta') = \exp\{-\|\eta - \eta'\|^2_2\}$. 
  
  
  
  After generating the covariates $X_i$, we consider a continuous exposure $A_i = r_a(X_i) + \nu_i$ and a continuous outcome $Y_i(a) = r_y(X_i) + a \, \gamma + \epsilon_i(a)$ with $\epsilon_i(a), \nu_i \iid \Normal(0,1)$ where $r_y(x) = r^\star_y(x) + r_a(x)$. We then generate $r^\star_y$ and $r_a$ as independent Gaussian processes with kernel $\rho(x,x') = \exp\{-\|x - x'\|^2_2\}$. Our parameter of interest is $\gamma$, which represents the causal effect of the exposure on the outcome. We consider two priors.


  \begin{description}[style=unboxed,leftmargin=0cm]
    \item[Naive] We impose \ref{ig3}, but otherwise use the ``true'' prior for $r^\star(x)$ using the kernel $2 \rho(x,x')$. We specify a $\Normal(0,10^2)$ prior for $\gamma$.
    \item[Direct] We use the model $Y_i(a) = r_y(X_i) + \omega \, \widehat r_a(X_i) + \gamma \, a$ where $\widehat r_a(x)$ is a pilot estimate of $r_a(x)$ obtained from fitting a Gaussian process to the relationship $A_i = r_a(X_i) + \nu_i$. We specify a $\Normal(0,10^2)$ prior for both $\gamma$ and $\omega$.
  \end{description}
  
  For the ground truth, we set $\gamma = 1$, $L = 1$, and consider $P \in \{10,200\}$, $N = 300$, and $\sigma_x = 2^{-j}$ where the $j$'s are evenly spaced between $-7$ and $-2$. 
  For each $\sigma_x$ and $P$ we generated $200$ simulated datasets and applied the Direct and Naive methods to estimate $\gamma$ and construct a 95\% credible interval. 
  
  \begin{figure}
    \centering
    \includegraphics[height=0.87\textheight]{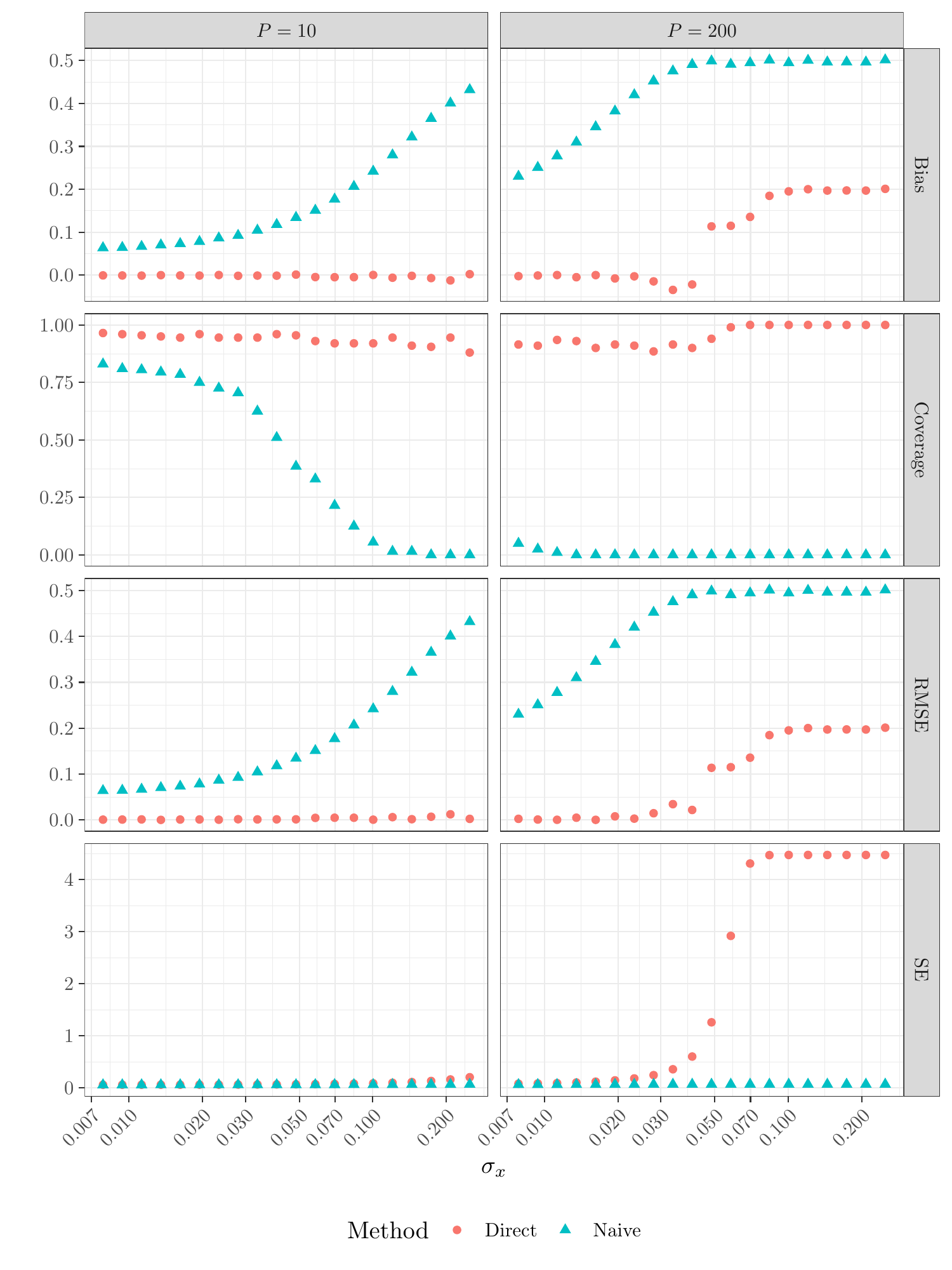}
    \caption{Results for the semiparametric regression on a manifold problem of Section~\ref{sec:when}. Bias denotes the average bias of $\widehat \gamma$, coverage denotes the coverage of nominal 95\% intervals, RMSE denotes the root-mean-squared-error in estimating $\gamma$, and SE denotes the average posterior standard deviation of $\gamma$.}
    \label{fig:sim_gp}
  \end{figure}

  Results are summarized in Figure~\ref{fig:sim_gp}, which displays the bias, coverage, RMSE, and average standard error. For $P = 10$, we see similar behavior as we did with the ridge regression problem: the Direct approach performs uniformly well and the naive approach performs much better as $\sigma_x$ is decreased. While the naive approach never does catch up to the direct approach, we do see that its deficiencies are attenuated as the $X_i$'s are generated closer and closer to $\sM$. For $P = 200$, while the naive approach does perform better as $\sigma_x$ is decreased, the problem appears to be too difficult for methods which do not explicitly account for dogmatism. The behavior of the direct approach is now very interesting, however. We note a sharp phase transition around $\sigma_x = 0.07$ where the problem essentially goes from infeasible to feasible --- the bias, RMSE, and standard error all decrease dramatically at this point. We also see that the direct approach is much more honest in terms of its uncertainty: when the problem is infeasible, the model correctly gives a large posterior standard error, whereas the naive model is always overconfident about its predictions.
  
\section{Discussion}

The main concrete recommendation we make in this article is that, for both causal inference and missing data problems, Bayesian ignorability (and in particular \ref{ig3}) corresponds to an informative prior on the degree of selection bias. This causes selection bias parameters to be regularized towards $0$, introducing substantial bias in high-dimensional or nonparametric problems and should not be imposed in most situations. Instead, Bayesians should reject \ref{ig3} by default in favor of a prior which allows for more direct control over the selection bias, and we have illustrated how to do this in several problems of interest. Of secondary interest, we have noted that certain features of the design can mitigate prior dogmatism about the selection bias, and showed that both ridge regression priors and Gaussian process priors possess some degree of adaptivity towards low-dimensional structures in $X_i$. But this does not change our general recommendation, as we consistently have observed improved performance of priors which reject \ref{ig3} even when such low-dimensional structures exist.


Dogmatism about other features of the model may also have deleterious effects on our inferences. In future work, we will extend our results to other problems in causal inference. Two of potential interest are estimation of the conditional average treatment effect (CATE) in observational studies and estimation of the natural direct and indirect effects in mediation analysis. In the latter setting, one must control for two different selection mechanisms: the effect of the confounders both on the treatment received and on the mediating variable.

While we have presented a number of corrections for dogmatism, we have not presented any coherent framework for \emph{deriving} corrections. This presents an important question: are there any objective Bayes principles which automatically lead to priors which adequately account for dogmatism? Certain strategies, such as using Jeffreys priors, cannot work because they usually \emph{imply} that \ref{ig3} holds. By contrast, other objective principles which are not parameterization invariant and do not necessarily imply \ref{ig3}, such as priors constructed from decision theoretic principles, entropy maximization, and reference priors have some chance of working \citep[see][for a review]{kass1996selection}. With the exception of entropy-maximizing priors, the computational difficulty of implementing these priors makes numeric experimentation difficult. Interestingly, entropy maximization with respect to the distribution of the observed data can be used to generate models which possess very strong Frequentist properties, but these models are (in our opinion) lacking a satisfying justification.

\bibliographystyle{apalike}
\bibliography{mybib.bib}

\singlespacing



\section*{Supplementary material}

# In Nonparametric and High-Dimensional Models, Ignorability is an Informative Prior

Throughout this supplement, we use the notation $X_P \asymp Y_P$ to mean that there exists a positive universal constant $C$ such that $C^{-1} Y_P \leq X_P \leq C Y_P$ for sufficiently large $P$ (the limiting variable $P$ usually being known from context). The notation $X_P = O_p(Y_P)$ means that for any $\delta$ we can find $K$ large enough that $\limsup_P \Pr(X < KY) \geq 1 - \delta$; again, the variable $P$ will generally be understood from context. The constants $C$ and $K$ are universal, allowed only to depend on the prior (but not on $P$). The statement $X \overset{\bullet}{\sim} \text{Normal}(a, b)$ means that $\frac{X-a}{b} \to \text{Normal}(0, 1)$ in distribution as $P \to \infty$.

## S.1 Proof of Proposition 1

We first derive $\Delta$ for the linear models. We note that $\mathbb{E}_\theta\{Y_i(a)\} = \gamma a$ while $\mathbb{E}_\theta\{Y_i(a) \mid A_i = a\} = \bar{X}(a)^\top \beta + \gamma a$ by ignorability, where $\bar{X}(a) = \mathbb{E}_\theta(X_i \mid A_i = a)$. Using the fact that $(X_i, A_i)$ is multivariate normal we have

$$\bar{X}(a) = \frac{\text{Cov}_\theta(X_i, A_i)}{\text{Var}_\theta(A_i)}\{a - \mathbb{E}_\theta(A_i)\} = a \frac{\text{Cov}(X_i, \phi^\top X_i)}{\sigma_a^2 + \text{Var}_\theta(\phi^\top X_i)} = a \frac{\Sigma \phi}{\sigma_a^2 + \phi^\top \Sigma \phi}.$$

Therefore

$$\mathbb{E}_\theta\{Y_i(a) \mid A_i = a\} - \mathbb{E}_\theta\{Y_i(a)\} = \bar{X}(a)^\top \beta = a \frac{\phi^\top \Sigma \beta}{\sigma_a^2 + \phi^\top \Sigma \phi} = a \frac{\sum_j \lambda_j W_j Z_j}{\sigma_a^2 + \sum_j \lambda_j Z_j^2},$$

where the last equality follows from the spectral decomposition $\Sigma = \Gamma \Lambda \Gamma^\top$.

For the semiparametric missing data problem we have

$$\mathbb{E}_\theta(Y_i \mid A_i = 1) - \mathbb{E}_\theta(Y_i) = \frac{\int \phi(x)\, y\, f_\beta(y \mid x)\, dy\, F_X(dx)}{\int \phi(x)\, F_X(dx)} - \int y\, f_\beta(y \mid x)\, dy\, F_X(dx)$$

$$= \frac{\int \phi(x)\, \beta(x)\, F_X(dx)}{\int \phi(x)\, F_X(dx)} - \int \beta(x)\, F_X(dx)$$

$$= \frac{\int \phi(x)\, \beta(x)\, F_X(dx) - \int \phi(x)\, F_X(dx) \int \beta(x)\, F_X(dx)}{\int \phi(x)\, F_X(dx)}$$

$$= \frac{\mathrm{Cov}_\theta\{\phi(X_i), \beta(X_i)\}}{\mathbb{E}_\theta\{\phi(X_i)\}}.$$

## S.2 Proof of Proposition 2

Define $V_j = \lambda_j W_j Z_j$ where $W$ and $Z$ are defined as in Proposition 1 and observe that $W_j \stackrel{\text{iid}}{\sim} \mathrm{Normal}(0, \tau_\beta^2)$ and $Z_j \stackrel{\text{iid}}{\sim} \mathrm{Normal}(0, \tau_\phi^2)$ because $\Gamma$ is an orthogonal transformation. Note that $\mathbb{E}(V_j) = 0$ and $\mathrm{Var}(V_j) = \lambda_j^2$. Define $s_P = \{\sum_j \mathrm{Var}(V_j)\}^{1/2} = \tau_\phi^2 \tau_\beta^2 \|\lambda\|_2$ where $\lambda = (\lambda_1, \ldots, \lambda_P)^\top$. By the Lyapounov central limit theorem (Billingsley, 1995, page 362) we have

$$\frac{\sum_j V_j}{s_P} \xrightarrow{d} \mathrm{Normal}(0,1) \quad \text{if} \quad \lim_{P \to \infty} s_P^{-(2+\epsilon)} \sum_{j=1}^P \mathbb{E}_\theta |V_j|^{2+\epsilon} = 0.$$

First, $\mathbb{E}_\theta |V_j|^{2+\epsilon} = C^2 \lambda_j^{2+\epsilon}$ where $C$ is the $2+\epsilon$ moment of the standard normal variable. Using the fact that $P^{-1} \sum_{j=1}^P \lambda_j^{2+\epsilon}$ converges, we have $\sum_{j=1}^P C^2 \lambda_j^{2+\epsilon} \asymp P$. Next, using the fact that $P^{-1} \sum_{j=1}^P \lambda_j^2 = \|\lambda\|_2^2 / P$ converges, we have $s_P \asymp P^{1/2}$. Hence $s_P^{-(2+\epsilon)} \sum_{j=1}^P \mathbb{E}_\theta |V_j|^{2+\epsilon} \asymp P^{-\epsilon/2}$, which therefore tends to 0 as $P \to \infty$. We conclude that $\sum_j V_j / s_P \xrightarrow{d} \mathrm{Normal}(0,1)$.

Next, note that

$$\mathbb{E}_\theta\Big(P^{-1} \sum_{j=1}^P \lambda_j Z_j^2\Big) = P^{-1} \tau_\phi^2 \sum_{j=1}^P \lambda_j \to \tau_\phi^2 \widetilde{\lambda} \quad \text{and}$$

$$\mathrm{Var}_\theta\Big(P^{-1} \sum_{j=1}^P \lambda_j Z_j^2\Big) = P^{-2} \mathrm{Var}(Z_1) \sum_{j=1}^P \lambda_j^2 \to 0.$$

Hence, by Chebychev's inequality, we have $\sum_j \lambda_j Z_j^2 / P \to \tau_\phi^2 \widetilde{\lambda}$ in probability. This gives

$$\frac{\sum_j V_j}{s_P \{\sigma_a^2 + \sum_j \lambda_j Z_j^2\}/P} \xrightarrow{d} \text{Normal}\left(0, \frac{1}{\tau_\phi^4 \widetilde{\lambda}^2}\right).$$

Equivalently, $\dfrac{\sum_j V_j}{\sigma_a^2 + \sum_j \lambda_j Z_j} \overset{\cdot}{\sim} \text{Normal}\left(0, \dfrac{s_P^2/P^2}{\tau_\phi^4 \widetilde{\lambda}^2}\right)$. But $s_P^2/P \to \tau_\phi^2 \tau_\beta^2 \bar{\lambda}^2$ by assumption. By Slutsky's theorem, this gives

$$\frac{\sum_j V_j}{\sigma_a^2 + \sum_j \lambda_j Z_j^2} \overset{\cdot}{\sim} \text{Normal}\left(0, \frac{\tau_\beta^2 \bar{\lambda}^2}{P \tau_\phi^2 \widetilde{\lambda}^2}\right).$$

Finally, multiplying both sides by $a$ gives the result.

## S.3 Proof of Proposition 3

From Proposition 1, we have

$$\Delta(1) = \frac{\phi^\top \Sigma b}{1 + \phi^\top \Sigma \phi} + \omega_0 \frac{\phi^\top \Sigma \phi}{1 + \phi^\top \Sigma \phi},$$

where $b \sim \text{Normal}(0, \tau^2 \mathrm{I}/P)$. From Proposition 2 we know the first term converges in probability to 0. It suffices to show that $\phi^\top \Sigma \phi \to \tau^2 \widetilde{\lambda}$ in probability. Defining $Z_j$ as in Proposition 1 we can rewrite this as $P^{-1} \sum_{j=1}^P \lambda_j Z_j^2$, which converges to $\tau^2 \widetilde{\lambda}$ in probability by Chebychev's inequality (c.f. the proof of Proposition 2).

## S.4 Proof Sketch for Theorem 1

We begin by reviewing some basic facts from linear algebra and random matrix theory. First, we recall the formula for inverting a block $2 \times 2$ matrix

$$\begin{pmatrix} A & B \\ C & D \end{pmatrix}^{-1} = \begin{pmatrix} (A - BD^{-1}C)^{-1} & -(A - BD^{-1}C)^{-1}BD^{-1} \\ -D^{-1}C(A - BD^{-1}C)^{-1} & D^{-1} + D^{-1}C(A - BD^{-1}C)^{-1}BD^{-1} \end{pmatrix}.$$

We also recall the empirical spectral distribution $G(dx)$ associated to the sample covariance matrix $S = \boldsymbol{X}^\top \boldsymbol{X}/N$, and let $m(z) = \int \frac{G(dx)}{x-z}$ denote the Stieltjes transform of $G(dx)$. Under HDA we have $P^{-1} \operatorname{tr}\{(S + \lambda \mathrm{I})^{-1}\} \to m(-\lambda)$ almost-surely. Letting $V = (S + \lambda \mathrm{I})^{-1}$ we also note for posterity the recursive identity $V^j S^k = V^{j-1} S^{k-1} - \lambda V^j S^{k-1}$, which implies in particular that $\widehat{\psi}_{jk} = \widehat{\psi}_{j-1,k-1} - \lambda \widehat{\psi}_{j,k-1}$ where $\widehat{\psi}_{jk} = \operatorname{tr}(V^j S^k)$. Moreover, $V$ and $S$ have the same eigenvectors and therefore commute.

The Bayes estimate of $(\gamma, \beta^\top)^\top$ under the conditions of Theorem 1 can be written as $M[\boldsymbol{A}, \boldsymbol{X}]^\top \boldsymbol{Y}$ where

$$M = \begin{pmatrix} \boldsymbol{A}^\top \boldsymbol{A} & \boldsymbol{A}^\top \boldsymbol{X} \\ \boldsymbol{X}^\top \boldsymbol{A} & \boldsymbol{X}^\top \boldsymbol{X} + N \lambda \mathrm{I} \end{pmatrix}^{-1}.$$

Applying the formula for inverting $2 \times 2$ block matricies, we have

$$\widetilde{\gamma} = \frac{\boldsymbol{A}^\top (\mathrm{I} - \boldsymbol{X} \bar{V} \boldsymbol{X}^\top) \boldsymbol{Y}/P}{\boldsymbol{A}^\top (\mathrm{I} - \boldsymbol{X} \bar{V} \boldsymbol{X}^\top) \boldsymbol{A}/P},$$

where $\bar{V} = (\boldsymbol{X}^\top \boldsymbol{X} + N \lambda \mathrm{I})^{-1}$. Write $\boldsymbol{Y} = \boldsymbol{X} b + \omega_0 \boldsymbol{X} \phi + \gamma_0 \boldsymbol{A} + \epsilon$ where $\epsilon \sim \operatorname{Normal}(0, \mathrm{I})$ independently of all other quantities. Reorganizing terms gives

$$\widetilde{\gamma} - \gamma_0 = \frac{\boldsymbol{A}^\top (\mathrm{I} - \boldsymbol{X} \bar{V} \boldsymbol{X}^\top)(\boldsymbol{X} b + \omega_0 \boldsymbol{X} \phi + \epsilon)/P}{\boldsymbol{A}^\top (\mathrm{I} - \boldsymbol{X} \bar{V} \boldsymbol{X}^\top) \boldsymbol{A}/P}$$

Conditioning on $(\boldsymbol{X}, \boldsymbol{A})$, iterated expectation gives

$$\mathbb{E}(\widetilde{\gamma} - \gamma_0) = \omega_0 \mathbb{E}\left(\frac{\boldsymbol{A}^\top (\mathrm{I} - \boldsymbol{X}\bar{V}\boldsymbol{X}^\top) \boldsymbol{X}\widetilde{\phi}/P}{\boldsymbol{A}^\top (\mathrm{I} - \boldsymbol{X}\bar{V}\boldsymbol{X}^\top) \boldsymbol{A}/P}\right) \tag{S.1}$$

where $\widetilde{\phi} = \bar{K}\boldsymbol{X}^\top \boldsymbol{A}$ is the BLUP of $\widetilde{\phi}$ and $\bar{K} = (\boldsymbol{X}^\top \boldsymbol{X} + N\eta \mathrm{I})^{-1}$. It can be shown that both the numerator and denominator of (S.1) converge both almost-surely and in $\mathscr{L}_1$ to a constant; taking this for granted, we compute the limiting value of the expectation of both expressions to derive the limits of the numerator and denominator in $\mathscr{L}_1$. Convergence of the above expectation to its associated limit then follows by bounded convergence because

$$\boldsymbol{A}^\top (\mathrm{I} - \boldsymbol{X}\bar{V}\boldsymbol{X}^\top)(\boldsymbol{A} - \boldsymbol{X}\widetilde{\phi}) = \boldsymbol{A}^\top (\mathrm{I} - \boldsymbol{X}\bar{V}\boldsymbol{X}^\top)(\mathrm{I} - \boldsymbol{X}\bar{K}\boldsymbol{X}^\top)\boldsymbol{A} \geq 0,$$

$$\boldsymbol{A}^\top (\mathrm{I} - \boldsymbol{X}\bar{V}\boldsymbol{X}^\top) \boldsymbol{X}\widetilde{\phi} = \boldsymbol{A}^\top (\mathrm{I} - \boldsymbol{X}\bar{V}\boldsymbol{X}^\top) \boldsymbol{X}\bar{K}\boldsymbol{X}^\top \boldsymbol{A} \geq 0$$

where non-negativity follows from the fact that $(\mathrm{I} - \boldsymbol{X}\bar{V}\boldsymbol{X}^\top)(\mathrm{I} - \boldsymbol{X}\bar{K}\boldsymbol{X}^\top)$ and $(\mathrm{I} - \boldsymbol{X}\bar{V}\boldsymbol{X}^\top)\boldsymbol{X}\bar{K}\boldsymbol{X}^\top$ are positive semi-definite (which can be quickly checked by taking the singular value decomposition of $\boldsymbol{X}$).

We begin by computing the denominator of (S.1). Noting that $\boldsymbol{A} \sim \mathrm{Normal}(0, \mathrm{I} + \boldsymbol{X}\boldsymbol{X}^\top/(N\eta))$, the expectation (conditional on $\boldsymbol{X}$) of the denominator is given by

$$\mathrm{tr}\left(\mathrm{I} + \frac{\boldsymbol{X}\boldsymbol{X}^\top}{N\eta} - \boldsymbol{X}\bar{V}\boldsymbol{X}^\top - \frac{\boldsymbol{X}\bar{V}\boldsymbol{X}^\top \boldsymbol{X}\boldsymbol{X}^\top}{N\eta}\right)/P = \left(N + \frac{\mathrm{tr}(S)}{\eta} - \mathrm{tr}(VS) - \frac{\mathrm{tr}(VS^2)}{\eta}\right)/P$$

$$= \frac{N + \widehat{\psi}_{01}/\eta - \widehat{\psi}_{11} - \widehat{\psi}_{12}/\eta}{P}.$$

Noting that $\widehat{\psi}_{00} = P$, several applications of the recursive identity for the $\widehat{\psi}$'s gives

$$\frac{N - P + \lambda\widehat{\psi}_{10} + (\lambda/\eta)(P - \lambda\widehat{\psi}_{10})}{P} \to \frac{1-r}{r} + \lambda m(-\lambda) + \frac{\lambda}{\eta}(1 - \lambda m(-\lambda))$$

using the fact that $\operatorname{tr}(V)/P = \widehat{\psi}_{10}/P \to m(-\lambda)$. Finally, we use the identity

$$1 - \lambda m(-\lambda) = \frac{1 - \lambda v(-\lambda)}{r}$$

to simplify this expression to $\frac{\lambda}{\eta r}[1 - (\lambda - \eta)v(-\lambda)]$. This latter identity follows from the fact that the spectral distributions of $\boldsymbol{X}\boldsymbol{X}^\top/N$ and $\boldsymbol{X}^\top\boldsymbol{X}/N$ differ only by $|N - P|$ zero eigenvalues, and is stated (for example) by Dobriban and Wager (2018).

Next, we compute the numerator of (S.1). Again using the distribution of $\boldsymbol{A}$ conditional on $\boldsymbol{X}$, the conditional expectation of the numerator is

$$\operatorname{tr}\left(\boldsymbol{X}\bar{K}\boldsymbol{X}^\top + \frac{\boldsymbol{X}\bar{K}\boldsymbol{X}^\top\boldsymbol{X}\boldsymbol{X}^\top}{N\eta} - \boldsymbol{X}\bar{V}\boldsymbol{X}^\top\boldsymbol{X}\bar{K}\boldsymbol{X}^\top - \frac{\boldsymbol{X}\bar{V}\boldsymbol{X}^\top\boldsymbol{X}\bar{K}\boldsymbol{X}^\top\boldsymbol{X}\boldsymbol{X}^\top}{N\eta}\right)/P$$
$$= \operatorname{tr}(SK + S^2K/\eta - S^2KV - S^3KV/\eta)/P$$

where $K = (S + \eta \mathrm{I})^{-1}$. Using the fact that $SV = \mathrm{I} - \lambda V$ and the related fact that $SK = \mathrm{I} - \eta V$, this expression simplifies to

$$\frac{\lambda}{\eta P}\operatorname{tr}(\mathrm{I} - \lambda V) \to \frac{\lambda}{\eta}(1 - \lambda m(-\lambda)) = \frac{\lambda}{\eta r}(1 - \lambda v(-\lambda)).$$

Finally, taking the ratio of limits of the numerator and denominator of (S.1) gives

$$\frac{1 - \lambda v(-\lambda)}{1 - (\lambda - \eta)v(-\lambda)}$$

as desired. The alternate form of the bias

$$\frac{\omega_0 \int x/(x + \lambda) \; F(dx)}{\int (x + \eta)/(x + \lambda) \; F(dx)}$$

follows from the definition of $v(-\lambda)$ as $\int F(dx)/(x + \lambda)$.

## S.5  Proof of Proposition 4

The expression $\Delta = \|\bar{\beta}\| \, \|\bar{\phi}\|/\mathbb{E}_\theta\{\phi(X_i)\} \langle \widetilde{\beta}, \widetilde{\phi} \rangle$ follows from Proposition 1. It suffices to show that $\|\bar{\beta}\| \, \|\bar{\phi}\|/\mathbb{E}_\theta\{\phi(X_i)\} = O_p(1)$. Since $\mathbb{E}_\theta\{\phi(X_i)\} \geq \delta$ and $\phi : \mathbb{R}^P \to [0,1]$, we only need to show $\|\bar{\beta}\| = O_p(1)$. By Markov's inequality and iterated expectation

$$\sup_P \Pr(\|\bar{\beta}\| > K) \leq \frac{\sup_P \mathbb{E}(\|\beta\|^2)}{K^2} = \frac{\sup_P \mathbb{E}\{\mathbb{E}_\theta \beta(X_i)^2\}}{K^2} = \frac{\sup_P \mathbb{E}\{\beta(X_i)^2\}}{K^2}$$

which can be made arbitrarily small by taking $K$ sufficiently large. Hence $\|\bar{\beta}\| = O_p(1)$.

### S.5.1  Proof of Proposition 5

A defining property of Gaussian processes on the Hilbert space $\mathscr{L}_2(F_X)$ is that if $\beta \sim \mathrm{GP}(0, K)$ then for any $\phi \in \mathscr{L}_2(F_X)$ we have $\langle \beta, \phi \rangle \sim \mathrm{Normal}(0, v)$ where the variance is given by $v = \iint \phi(x) \, \phi(x') \, K(x,x') \, F_X(dx)$ (van Zanten and van der Vaart, 2008). The first equality of Proposition 5 is immediate from this fact because $\Delta = \langle \bar{\beta}, \bar{\phi} \rangle / \mathbb{E}\{\phi(X_i)\}$ and $\bar{\beta} \sim \mathrm{GP}(0, \tau_\beta^2 \, \bar{\rho})$.

For the second equality, let $\rho(x,x') = \sum_{j=1}^\infty \lambda_j \, v_j(x) \, v_j(x')$ be the Karhunen–Loève expansion of $\rho(x,x')$, i.e., the $v_j$'s are an orthonormal system in $\mathscr{L}_2(F_X)$ and $\lambda_1 \geq \lambda_2 \geq \cdots$. Then we can write $\beta(x) = \tau_\beta \sum_{j=1}^\infty \beta_j \, v_j(x)$ where $\beta_j \sim \mathrm{Normal}(0, \lambda_j)$ (see, e.g., Alexanderian, 2015). Centering $\beta$ gives $\bar{\beta}(x) = \tau_\beta \sum_j \beta_j \, \bar{v}_j(x)$ where $\bar{v}_j = v_j - \int v_j \, dF_X$, and it is easy to show that the $\bar{v}_j$'s are also an orthogonal system. From fact that $\{\bar{v}_j : j = 1, 2, \ldots\}$ is an orthogonal system we have

$$\mathrm{Cov}_\theta\{\beta(X_i), \phi(X_i)\} = \langle \bar{\beta}, \bar{\phi} \rangle = \tau_\beta \sum_{j=1}^\infty \beta_j \, \langle \bar{v}_j, \bar{\phi} \rangle = \tau_\beta \sum_{j=1}^\infty \beta_j \, \mathrm{Cov}\{v_j(X_i), \phi(X_i)\}.$$

Hence $\mathrm{Var}[\mathrm{Cov}_\theta\{\beta(X_i), \phi(X_i)\}] = \tau_\beta^2 \sum_{j=1}^\infty \lambda_j \, \mathrm{Cov}\{v_j(X_i), \phi(X_i)\}^2$, using the fact that if $S_n = \sum_{j=1}^n T_j$ is a random sum of uncorrelated random variables such that $\sum_{j=1}^n \mathbb{E}(T_j)$ and $\sum_{j=1}^n \mathrm{Var}(T_j)$ both converge then $\sum_{j=1}^\infty T_j$ converges almost-surely with $\mathrm{Var} \sum_{j=1}^\infty T_j =$

$\sum_{j=1}^{\infty} \text{Var}(T_j)$.

## S.5.2 Proof of Proposition 6

We begin by computing $\Omega = \iint \rho(x, x') \, \text{Normal}(x \mid 0, \Sigma) \, \text{Normal}(x' \mid 0, \Sigma) \, dx \, dx'$. Note that $\rho(x, x') = \det(2\pi H)^{1/2} \, \text{Normal}(x \mid x', H)$. We will use the standard fact about multivariate normal distributions that for any vectors $z, x, w$ and covariance matricies $A$ and $B$ we have $\text{Normal}(z \mid x, A) \, \text{Normal}(x \mid w, B) = \text{Normal}(z \mid w, A+B) \, \text{Normal}(x \mid m, S)$ for some $m$ and $S$; hence $\int \text{Normal}(z \mid x, A) \, \text{Normal}(x \mid w, B) \, dx = \text{Normal}(z \mid w, A+B)$. Using this fact, we can write $\text{Normal}(0 \mid x, \Sigma) \, \text{Normal}(x \mid x', H) = \text{Normal}(0 \mid x', \Sigma + H) \, \text{Normal}(x \mid m, S)$ for some vector $m$ and covariance $S$. Integrating out $x$ gives

$$\Omega = \det(2\pi H)^{1/2} \int \text{Normal}(0 \mid x', \Sigma + H) \, \text{Normal}(x' \mid 0, \Sigma) \, dx'.$$

Using this basic fact again we have $\text{Normal}(0 \mid x', \Sigma + H) \, \text{Normal}(x' \mid 0, \Sigma) = \text{Normal}(0 \mid 0, 2\Sigma + H) \, \text{Normal}(x' \mid m, S)$ for some $(m, S)$ so that

$$\Omega = \det(2\pi H)^{1/2} \, \text{Normal}(0 \mid 0, 2\Sigma + H) = \sqrt{\frac{\det(H)}{\det(H + 2\Sigma)}}.$$

By the triangle inequality, note that $|\bar{\rho}(x, x')| \le 4\rho(x, x')$. By the crude upper bound $|\bar{\phi}(x)| \le 1$ we have that the constant $c$ from Proposition 5 satisfies

$$\begin{aligned} c &\le \frac{\tau_\beta^2}{\mathbb{E}\{\phi(X_i)\}^2} \iint \bar{\phi}(x) \, \bar{\phi}(x') \, |\bar{\rho}(x, x')| \, F_X(dx) \, F_X(dx') \\ &\le \frac{4\tau_\beta^2 \, \Omega}{\mathbb{E}\{\phi(X_i)\}^2}. \end{aligned}$$

All that remains is to show that $\det(H)/\det(H + 2\Sigma)$ decays exponentially. To do this we apply Minkowski's determinant theorem, which states that if $A$ and $B$ are positive semidefinite matrices then $\det(A + B)^{1/P} \ge \det(A)^{1/P} + \det(B)^{1/P}$ (Marcus and Minc, 1964,

| $(N = 250, P = 5)$ | Method | RMSE | Coverage | Estimated SE | Empirical SE |
|---|---|---|---|---|---|
| Nonlinear/Homogeneous | Naive | 0.24 | 0.94 | 0.24 | 0.22 |
|  | IPW | 0.46 | 0.95 | 0.30 | 0.45 |
|  | SOP | 0.26 | 0.98 | 0.31 | 0.26 |
|  | SOP-GP | 0.19 | 0.97 | 0.21 | 0.19 |
| Nonlinear/Heterogeneous | Naive | 0.25 | 0.94 | 0.25 | 0.24 |
|  | IPW | 0.45 | 0.96 | 0.32 | 0.44 |
|  | SOP | 0.32 | 0.97 | 0.36 | 0.32 |
|  | SOP-GP | 0.21 | 0.96 | 0.23 | 0.21 |
| Linear/Homogeneous | Naive | 0.19 | 0.94 | 0.18 | 0.19 |
|  | IPW | 0.25 | 0.94 | 0.24 | 0.25 |
|  | SOP | 0.20 | 0.95 | 0.20 | 0.20 |
|  | SOP-GP | 0.20 | 0.93 | 0.19 | 0.20 |
| Linear/Heterogeneous | Naive | 0.20 | 0.94 | 0.20 | 0.20 |
|  | IPW | 0.28 | 0.95 | 0.26 | 0.28 |
|  | SOP | 0.28 | 0.94 | 0.28 | 0.27 |
|  | SOP-GP | 0.22 | 0.94 | 0.22 | 0.22 |

Table S.1: Results for the simulation study of Section 4.3 when $N = 250$ and $P = 5$.

page 115). This gives

$$\frac{\det(H)}{\det(H + 2\Sigma)} \leq \frac{\det(H)}{(\det(H)^{1/P} + 2\det(\Sigma)^{1/P})^P} = \left(1 + 2\frac{\det(\Sigma)^{1/P}}{\det(H)^{1/P}}\right)^{-P}.$$

Because $\liminf_P \det(\Sigma)^{1/P}/\det(H)^{1/P} \equiv d > 0$, the left hand side is bounded by $(1+d)^{-P}$ for sufficiently large $P$, completing the proof of the general case.

The special cases (a) and (b) can both be easily checked by showing that $\det(\Sigma)^{1/P}/\det(H)^{1/P}$ is bounded away from 0.

## S.6 Additional Simulation Results for Section 4.3

Raw results of the simulation of Section 4.3 are given in Tables S.1, S.2, S.3, and S.4.

| $(N = 500, P = 5)$ | Method | RMSE | Coverage | Estimated SE | Empirical SE |
|---|---|---|---|---|---|
| Nonlinear/Homogeneous | Naive | 0.17 | 0.94 | 0.16 | 0.17 |
|  | IPW | 0.20 | 0.94 | 0.19 | 0.20 |
|  | SOP | 0.20 | 0.97 | 0.22 | 0.20 |
|  | SOP-GP | 0.14 | 0.94 | 0.14 | 0.14 |
| Nonlinear/Heterogeneous | Naive | 0.17 | 0.93 | 0.17 | 0.17 |
|  | IPW | 0.21 | 0.93 | 0.19 | 0.21 |
|  | SOP | 0.23 | 0.97 | 0.25 | 0.23 |
|  | SOP-GP | 0.16 | 0.93 | 0.16 | 0.15 |
| Linear/Homogeneous | Naive | 0.12 | 0.96 | 0.13 | 0.12 |
|  | IPW | 0.16 | 0.95 | 0.16 | 0.16 |
|  | SOP | 0.14 | 0.95 | 0.14 | 0.14 |
|  | SOP-GP | 0.12 | 0.96 | 0.13 | 0.12 |
| Linear/Heterogeneous | Naive | 0.14 | 0.96 | 0.14 | 0.13 |
|  | IPW | 0.17 | 0.95 | 0.17 | 0.17 |
|  | SOP | 0.18 | 0.97 | 0.19 | 0.18 |
|  | SOP-GP | 0.14 | 0.96 | 0.15 | 0.14 |

Table S.2: Results for the simulation study of Section 4.3 when $N = 500$ and $P = 5$.

| $(N = 250, P = 20)$ | Method | RMSE | Coverage | Estimated SE | Empirical SE |
|---|---|---|---|---|---|
| Nonlinear/Homogeneous | Naive | 4.35 | 0.01 | 0.44 | 2.54 |
|  | IPW | 0.67 | 0.94 | 0.60 | 0.65 |
|  | SOP | 0.30 | 0.95 | 0.31 | 0.29 |
|  | SOP-GP | 0.29 | 0.94 | 0.29 | 0.29 |
| Nonlinear/Heterogeneous | Naive | 4.72 | 0.01 | 0.48 | 2.58 |
|  | IPW | 0.70 | 0.94 | 0.64 | 0.66 |
|  | SOP | 0.35 | 0.95 | 0.36 | 0.35 |
|  | SOP-GP | 0.33 | 0.95 | 0.33 | 0.33 |
| Linear/Homogeneous | Naive | 1.91 | 0.24 | 0.24 | 1.38 |
|  | IPW | 0.44 | 0.95 | 0.34 | 0.44 |
|  | SOP | 0.22 | 0.94 | 0.20 | 0.22 |
|  | SOP-GP | 0.22 | 0.93 | 0.20 | 0.22 |
| Linear/Heterogeneous | Naive | 2.43 | 0.26 | 0.29 | 1.52 |
|  | IPW | 0.47 | 0.95 | 0.40 | 0.47 |
|  | SOP | 0.29 | 0.93 | 0.28 | 0.29 |
|  | SOP-GP | 0.28 | 0.93 | 0.26 | 0.28 |

Table S.3: Results for the simulation study of Section 4.3 when $N = 250$ and $P = 20$.

| $(N = 500, P = 20)$ | Method | RMSE | Coverage | Estimated SE | Empirical SE |
|---|---|---|---|---|---|
| Nonlinear/Homogeneous | Naive | 1.72 | 0.00 | 0.22 | 1.11 |
| | IPW | 0.36 | 0.96 | 0.36 | 0.36 |
| | SOP | 0.19 | 0.97 | 0.22 | 0.19 |
| | SOP-GP | 0.17 | 0.96 | 0.19 | 0.17 |
| Nonlinear/Heterogeneous | Naive | 1.81 | 0.00 | 0.23 | 1.17 |
| | IPW | 0.35 | 0.95 | 0.37 | 0.34 |
| | SOP | 0.22 | 0.97 | 0.25 | 0.22 |
| | SOP-GP | 0.20 | 0.94 | 0.22 | 0.20 |
| Linear/Homogeneous | Naive | 0.74 | 0.46 | 0.14 | 0.62 |
| | IPW | 0.33 | 0.93 | 0.23 | 0.33 |
| | SOP | 0.15 | 0.94 | 0.14 | 0.15 |
| | SOP-GP | 0.14 | 0.94 | 0.14 | 0.14 |
| Linear/Heterogeneous | Naive | 0.83 | 0.53 | 0.17 | 0.71 |
| | IPW | 0.34 | 0.95 | 0.26 | 0.33 |
| | SOP | 0.18 | 0.97 | 0.19 | 0.18 |
| | SOP-GP | 0.17 | 0.96 | 0.18 | 0.17 |

Table S.4: Results for the simulation study of Section 4.3 when $N = 500$ and $P = 20$.

## S.7 Numerical Verification of Bias Formula

Figure 1 gives numerical results verifying the accuracy of the bias formulas derived in Section 2.1 and Section 4.1. We see that both formulas are extremely accurate for $N = 500$ across a range of values of $(\lambda, \eta, r)$. We set $\Sigma = I$ in this figure.

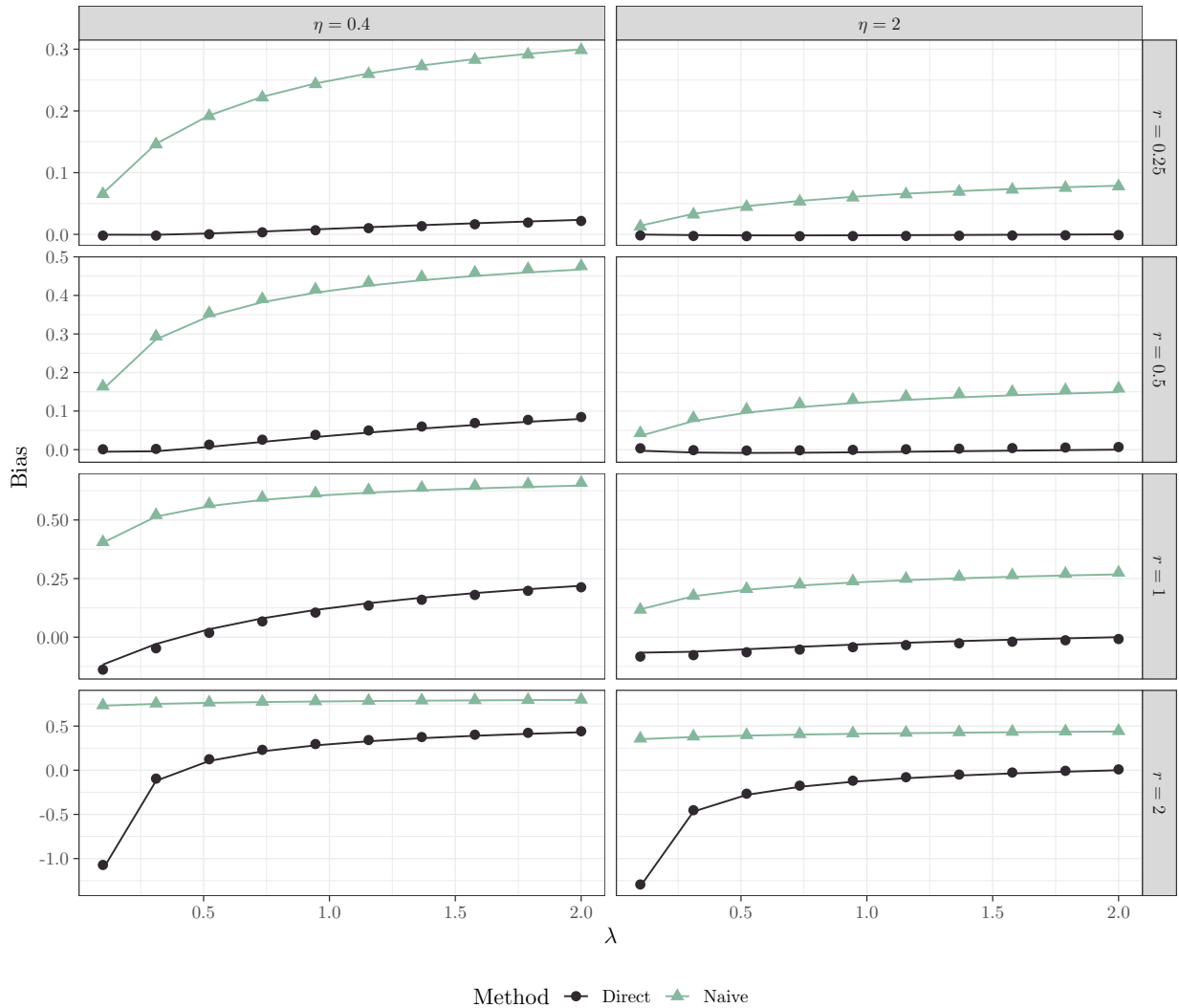

Figure 1: Comparison of the formulas for the asymptotic bias for the direct Z-prior and the naive ridge regression prior to their respective empirical bias for various values of $(\eta, r, \lambda)$ with $\omega_0 \equiv 1$, $\Sigma \equiv I$, and $N = 500$. The empirical bias was computed by fitting each method 200 times and computing $\widetilde{\gamma} - \gamma_0$.



\end{document}